\documentclass[prb,reprint,aps,superscriptaddress,longbibliography]{revtex4-2}
\usepackage{graphicx}
\usepackage{dcolumn}
\usepackage{bm}
\usepackage{braket}
\usepackage{amsmath}
\usepackage{amssymb}
\usepackage{feynmp}
\usepackage{graphics}
\usepackage{enumerate}
\usepackage[dvipsnames]{xcolor}
\usepackage{comment}
\usepackage[normalem]{ulem} 
\usepackage[hidelinks]{hyperref}

\begin{document}

\title{Magic Angles and Fractional Chern Insulators in Twisted Homobilayer TMDs}

\author{Nicol\'as Morales-Dur\'an}
\email{na.morales92@utexas.edu}
\affiliation{Department of Physics, The University of Texas at Austin, Austin, Texas, 78712, USA}

\author{Nemin Wei}
\affiliation{Department of Physics, The University of Texas at Austin, Austin, Texas, 78712, USA}

\author{Jingtian Shi}
\affiliation{Department of Physics, The University of Texas at Austin, Austin, Texas, 78712, USA}

\author{Allan H. MacDonald}
\affiliation{Department of Physics, The University of Texas at Austin, Austin, Texas, 78712, USA}

\begin{abstract}
We explain the appearance of magic angles and fractional Chern insulators
in twisted K-valley homobilayer transition metal dichalcogenides by mapping their continuum model 
to a Landau level problem.  
Our approach relies on an adiabatic approximation for the quantum mechanics of valence band 
holes in a layer-pseudospin field that is valid for sufficiently small twist angles
and on a lowest Landau level approximation that is valid for sufficiently large twist angles.
It simply explains why the quantum geometry of the lowest moir\'e miniband is nearly ideal at 
particular flat-band twist angles, predicts that topological flat bands occur only when the 
valley-dependent moir\'e potential is sufficiently strong compared to the interlayer tunneling amplitude, 
and provides a powerful starting point for the study of interactions.
\end{abstract} 

\maketitle

{\em Introduction}--- Recent experiments \cite{FCI_Experiment1,FCI_Experiment2,FCI_Transport1,FCI_Transport2,lu2023fractional} 
have reported the first observations of fractional Chern insulator (FCI) states, 
exotic states of matter that display a fractional quantum Hall effect in the 
absence of a magnetic field \footnote{Recently the term  
fractional Chern insulator has also been used \cite{xie2021fractional} 
to describe fractional quantum Hall states that occur
in a moiré band at non-zero magnetic field.  If this terminology is adopted, the zero-field states 
of interest here, which combine magnetism with fractionalization since they break time-reversal
spontaneously, should be referred to as fractional quantum anomalous Hall states.}.
It has been understood for some time \cite{Wen_FCI,DasSarma_Sun_FCI,Neupert_FCI,Bernevig_Regnault_FCI,Sheng_FCI,Parameswaran_Ideal}
that FCI states do occur in artificial theoretical model
systems. In this Letter we address FCI states in the hole fluids of AA-stacked $K$-valley transition metal 
dichalcogenide (TMD) twisted homobilayers, where the effect was first observed \cite{FCI_Experiment1,FCI_Experiment2}.
Earlier theoretical work had hinted that FCI states might 
appear in this type of two-dimensional electron system by   
showing that their moiré minibands could carry Chern numbers \cite{FengchengTopology,FengchengTopology2}, 
that the moir\'e band width could mysteriously vanish \cite{LiangFuMagic,KaiSun_FCI,Crepel_FCI} near a magic twist angle, 
and that the bands have almost ideal quantum geometry \cite{FCI_Flatiron} when flat \footnote{We do not specifically address the FQAH very recently observed \cite{lu2023fractional} in rhombohedral graphene stacks.}.
There are however many open questions; for example, the FCI states so far appear at a few hole filling fractions and they appear over a wider regime of twist angle than theoretically expected.  In this Letter we address the 
most baffling question - why do the magic angles appear in the first place? Our answer points to 
a strategy for quantitative descriptions of these moiré FCI states. 

Continuum models of TMD moirés \cite{wu2018hubbard,FengchengTopology} are expected to give an accurate description of their low-energy physics.
In bilayers, the layer-dependent terms can always be expressed in terms of an effective field that acts on the layer pseudospin.
For AA-stacked K-valley homobilayers \cite{FengchengTopology,FengchengTopology2,LiangFuMagic} the effective field has a topologically 
non-trivial spatial structure with one Skyrmion for each moir\'e period. It is natural to suspect that there is a connection between the real space Skyrmion lattice and the momentum space Chern numbers,
although it was recognized from the beginning \cite{FengchengTopology} that the correspondence is not universal. 
Instead the Chern number of the topmost valence moiré miniband depends on the phenomenological parameters
$(V_m,\psi,\omega)$ that enter the continuum model, whose values vary from system to system \cite{FengchengTopology,FengchengTopology2,FCI_Flatiron,LiangFuMagic,FCI_DiXiao,FCI_LiangFu},
and can vanish even though the Skyrmion lattice is always present. Here $V_m$, $\psi$, and $\omega$ respectively specify the strength and shape 
of the moir\'e potentials in each layer, and the strength of interlayer tunneling.  

\begin{figure*}
\centering
\includegraphics[width=0.8\linewidth]{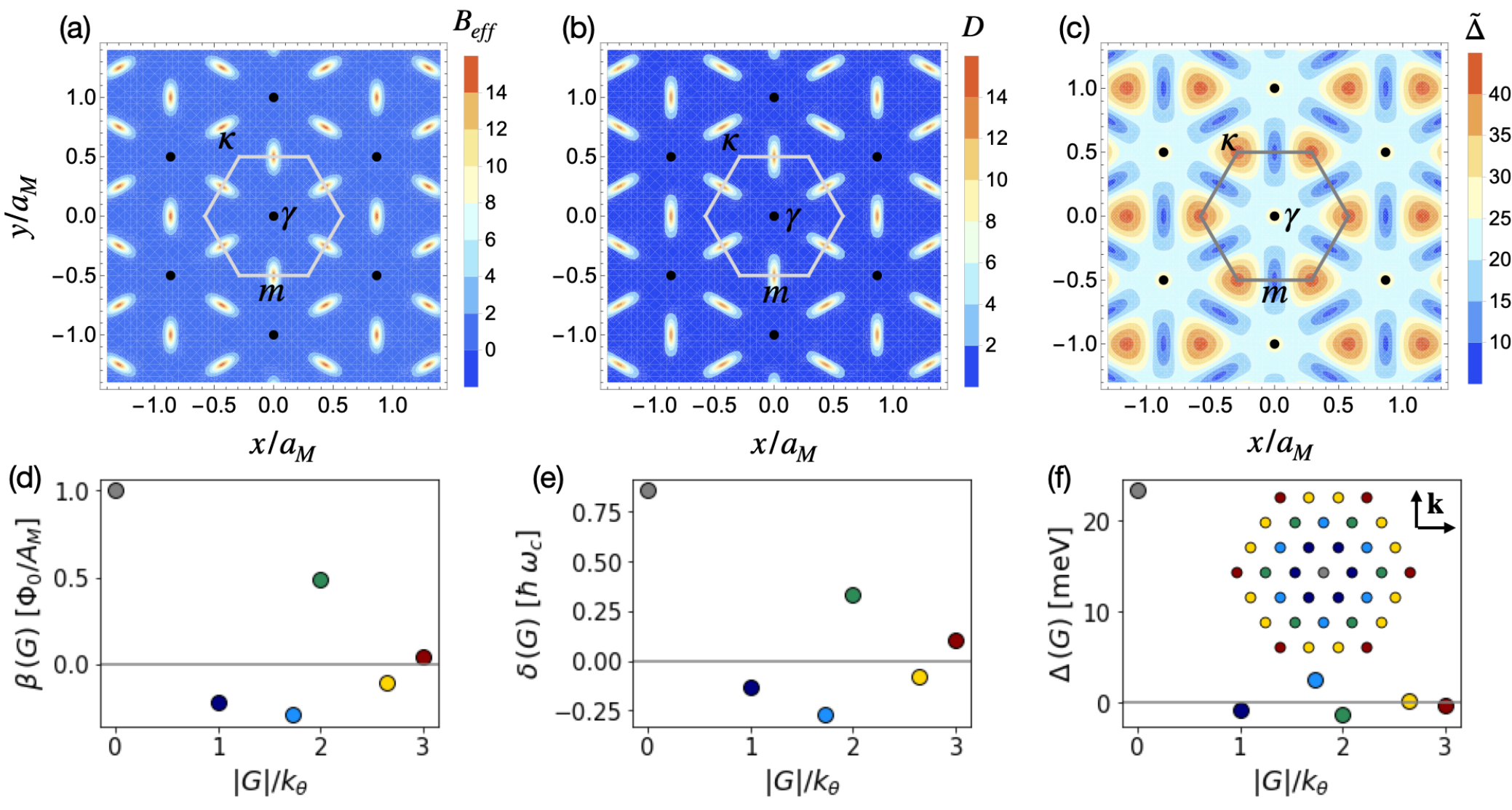}
\caption{Spatial distribution of the (a) effective magnetic 
field ${\bm B}_{\text{eff}}({\bm r})$ generated by the layer pseudospin Skyrmion in units of flux quantum per unit cell area,
(b) the kinetic potential $D$ in units of $\hbar\,\omega_c$, and (c) the effective Zeeman field $\tilde{\Delta}({\bm r})$ in meV. 
Black dots indicate moiré superlattice sites. The Wigner-Seitz cell boundary is marked by solid lines and the 
$\kappa$, $m$ and $\gamma$ high symmetry points that are key to magic angle behavior (see main text) are indicated. 
(d)-(f) The corresponding Fourier expansion coefficients.
The inset in (f) shows the first six shells of reciprocal lattice vectors. The magnetic form factors have the numerical values $1,0.163,4\times10^{-3},7\times10^{-4},3\times10^{-6},8\times10^{-8}$ for the six plotted shells. These illustrative plots 
are for unstrained MoTe$_2$ \cite{FengchengTopology} model parameters: $V_m=8$ meV, $\psi=89.6^{\circ}$ and $\omega=-8.5$ meV.}
\label{fig:Effective_MagField}
\end{figure*}

In this Letter we exploit an approximation to the TMD continuum model that is motivated by the 
presence of the Skyrmion lattice, 
one that maps it to holes in Landau levels subject to a periodic potential, to explain the magic angle behavior.
We start by using an adiabatic approximation for the layer pseudospin to transform the continuum Hamiltonian into one for 
layerless holes under the effect of a periodic potential and a periodic magnetic field with a non-zero mean. 
By separating the effective magnetic field into average and sinusoidal contributions, 
we further project the problem to the lowest Landau level (LLL) induced by the average
effective magnetic field, whose strength is one flux quantum per moir\'e unit cell. Within the LLL, both field and potential variations can be grouped into an effective potential with honeycomb lattice
symmetry that is accurately characterized by a single real parameter $\xi_1$, whose value
is determined by the continuum model parameters. We show that the magic angle behavior occurs 
when $\xi_1$ vanishes.  At the magic angle,
our series of transformations has mapped the bilayer Hamiltonian to 
the ordinary fractional quantum Hall problem, making the fractional Hall effect inevitable.

{\em Adiabatic Approximation}---
We start from the continuum model Hamiltonian for TMD homobilayers \cite{FengchengTopology}, 
\begin{align}
    H_{TMD}=-\frac{\hbar^2\,{\bm k}^2}{2m^*}\,\sigma_0+{\bm \Delta}({\bm r})\cdot {\bm \sigma}+\Delta_0({\bm r})\,\sigma_0,
    \label{eq:continuummodel}
\end{align}
where ${\bm \Delta}=(\text{Re}\,\Delta_T,\text{Im}\,\Delta_T,(\Delta_b-\Delta_t)/2)$,
$\Delta_T$ is the complex interlayer tunneling amplitude, $\Delta_0=(\Delta_b+\Delta_t)/2$, where $\Delta_t$ and 
$\Delta_b$ are the potential energies in the top and bottom layers, ${\bm \sigma}$ are the Pauli matrices and $\sigma_0$ the identity matrix. Eq.~\eqref{eq:continuummodel} is a valley-projected single-particle Hamiltonian; the full 
Hilbert space includes two valleys that are related to each other by time-reversal. For details on the continuum model see the supplemental material \cite{Supplemental}.

Next we apply a unitary transformation $U({\bm r})$ that rotates ${\bf \Delta}({\bm r})$ to the $z$-direction at 
each position \cite{Millis_TopologicalHall,Nagaosa_TopologicalHall_1,Nagaosa_TopologicalHall_2,van2013magnetic}
\begin{align}
    U^{\dagger}({\bm r})\left[ {\bm \Delta}({\bm r})\cdot {\bm \sigma}\right]U({\bm r})=|{\bm \Delta}({\bm r})|\sigma_z.
\end{align}
Because the transformation is position-dependent, the kinetic energy term 
includes coupling between the up and down pseudospin sectors.
Projection to the up pseudospin sector can, however, be justified 
when the ${\bm r}$-dependence is slow. 
After projection to the up pseudospin sector, which we will refer to as the adiabatic approximation, 
the matrix Hamiltonian operator reduces to a scalar.  Because of the real space Berry phases 
associated with the Skyrmion lattice \cite{FengchengTopology} in the pseudospin field, the
kinetic-energy operator gains an effective periodic magnetic field with non-zero mean. 
Additionally, there is a contribution from the off-diagonal part of the matrix Hamiltonian, the kinetic potential
$D=(\hbar^2/8m^*)\sum_{i=x,y}\left[\partial_i {\bm n}\right]^2$, which is the local increase 
in kinetic energy due to the position-dependence of the layer spinor, with ${\bm n}({\bm r})=\Delta(\bm{r})/|\Delta(\bm{r})|$. The effective Zeeman energy is 
$\tilde{ \Delta}=|{\bm \Delta}|+\Delta_0$, yielding \cite{volovik1987linear,bruno2004topological,Skyrmions_LiangFu,Skyrmions_TMD1,Skyrmions_TMD2} 
\begin{align}
    H=-\frac{1}{2m^*}\left[\hbar{\bm k}+e\tilde{{\bm A}}({\bm r})\right]^2 - D({\bm r})+\tilde{\Delta}({\bm r}).
\label{MagneticHamiltonian}
\end{align}
The adiabatic approximation is valid when $|{\bm \Delta}({\bm r})|\gg \hbar^2/(m^* A_M)$ where $A_M$ is the moir\'e unit cell area.
The emergent magnetic field in Eq.~\eqref{MagneticHamiltonian} is proportional \cite{FengchengTopology2} to the Pontryagin index density of ${\bm n}({\bm r})$,
\begin{align}
    {\bm B}_{\text{eff}}({\bm r})={\bm \nabla} \times \tilde{{\bm A}}({\bm r})=\frac{\hbar}{2e}\,{\bm n}\cdot \left( \partial_x {\bm n}\times \partial_y{\bm n} \right),
\end{align}
and therefore has one flux quantum per moir\'e period.  In magnetic thin films with non-collinear spin textures
a similar effective magnetic field is responsible for 
the topological Hall effect \cite{Millis_TopologicalHall,Nagaosa_TopologicalHall_1,Nagaosa_TopologicalHall_2,van2013magnetic}. See supplemental material for details on how to obtain Eq.~\eqref{MagneticHamiltonian} \cite{Supplemental}.

Fig. \ref{fig:Effective_MagField}(a) shows the spatial dependence of the effective magnetic field for continuum model parameters corresponding to unstrained MoTe$_2$. ${\bm B}_{\text{eff}}$ has three sharp peaks per period centered on the $m$ points of the Wigner-Seitz cell. We separate the effective magnetic field into an average value, ${\bm B}_0 = \Phi_0/A_M$, where $\Phi_0$ is the magnetic 
flux quantum, and a position-dependent part, denoted by ${\bm B}({\bm r})$, that has zero average.  The corresponding vector potential can be split in a similar way so that 
\begin{align}
    {\bm B}_{\text{eff}}({\bm r})={\bm B}_0+{\bm B}({\bm r})={\bm \nabla}\times{\bm A}_0+{\bm \nabla}\times{\bm{ A}}({\bm r});
\end{align}
${\bm A}_0$ is a linear function of position while ${\bm{ A}}({\bm r})$ has the moiré superlattice periodicity. The adiabatic Hamiltonian becomes 
\begin{align}
    H&=-\frac{\hbar^2}{2m^*}\left[ {\bm \Pi}+\frac{e}{\hbar}\,{\bm{ A}}({\bm r})\right]^2-D(\bm r)+\tilde{\Delta}({\bm r}),
    \label{HamiltonianAdiabatic}
\end{align}
where we have defined ${\bm \Pi}={\bm k}+e\,{\bm A}_0/\hbar$. The shape of $D$ in
Fig. \ref{fig:Effective_MagField}(b) is similar to that of ${\bm B}_{\text{eff}}$.
Both quantities are peaked near the $m$-points of the Wigner-Seitz cell, midway between 
the chalcogen on metal (XM) and metal on chalcogen points (MX) at the $\kappa$ Wigner-Seitz cell 
corners.
The spatial distribution of the effective Zeeman field 
$\tilde{ \Delta}$ is shown in Fig. \ref{fig:Effective_MagField}(c). The peaks at $\kappa$ are due to large potential difference between layers, whereas those at $\gamma$ are due to peaks 
in interlayer tunneling at metal on metal (MM) positions. As we will explain, the magic angle behavior is 
intimately related to the spatial pattern of the effective Zeeman field.

Because ${\bm B}_{\text{eff}}$, $D$ and $\tilde{\Delta}$ are periodic functions, they have the moir\'e 
lattice Fourier expansion 
\begin{eqnarray}
    {\bm B}_{\text{eff}}({\bm r})&=&\sum_{{\bm G}}\beta({\bm G})\,e^{i\, {\bm G}\cdot {\bm r}}, \\
    D(\bm r)&=& \sum_{{\bm G}}\delta({\bm G})\,e^{i\, {\bm G}\cdot {\bm r}},\\
    \tilde{\Delta}(\bm r) &=& \sum_{{\bm G}}\Delta({\bm G})\,e^{i\, {\bm G}\cdot {\bm r}},
\end{eqnarray}
where ${\bm G}$ are reciprocal lattice vectors. Since these three functions have $C_6$ rotational symmetry, the Fourier coefficients are identical within reciprocal 
lattice vector shells and real.  
Fig. \ref{fig:Effective_MagField}(d)-(f) shows the Fourier expansion coefficients for the first six shells of
${\bm B}_{\text{eff}}$, $D(\bm{r})$ and $\tilde{\Delta}$, respectively. The kinetic momentum term in Eq. \eqref{HamiltonianAdiabatic}  can then be expressed in terms of the Landau level
ladder operators $a$ and $a^{\dagger}$ and the complex vector potential $ A_{\pm}=A_x\pm iA_y$, see \cite{Supplemental}. Using $A({\bm k})=i{\bm k}\times {\bm B}({\bm k})/|{\bm k}|^2$, 
we find that $A_{\pm}({\bm G})=\sum_{{\bm G}}\alpha_{\pm}({\bm G})e^{i\, {\bm G}\cdot {\bm r}}$, with the Fourier coefficients given by
\begin{align}
    \alpha_{\pm}({\bm G})=\frac{\pm G_x+i\,G_y}{|{\bm G}|^2}\,\beta({\bm G}).
    \label{eq:alphapm}
\end{align}
It follows that the Landau level representation of the adiabatic Hamiltonian is 
\begin{widetext}
\begin{align}
    H=&-\hbar\omega_c\left( a^{\dagger}a+\frac{1}{2}\right)+\frac{i\,e\,\hbar}{\sqrt{2}\,m^*\ell}\sum_{\bm G}\left(a\,\alpha_+({\bm G})-a^{\dagger}\,\alpha_-({\bm G})\right)\,e^{i\, {\bm G}\cdot {\bm r}}\nonumber \\
    &-\frac{e^2}{2m^*}\, \sum_{{\bm G},{\bm G'}}\alpha_+({\bm G})\,\alpha_-({\bm G'})\,\,e^{i\, ({\bm G}+{\bm G'})\cdot {\bm r}}-\sum_{{\bm G}}\delta({\bm G})\,e^{i\, {\bm G}\cdot {\bm r}}+\sum_{{\bm G}}\Delta({\bm G})\,e^{i\, {\bm G}\cdot {\bm r}}.
    \label{Hamiltonian_alphamodes}
\end{align}
\end{widetext}
In Eq.~\eqref{Hamiltonian_alphamodes} $\alpha_{\pm}$ has been expressed in units of $\Phi_0/A_{M}$,
$\hbar \omega_c = 2 \pi \hbar^2/(m^* A_M) \approx 2.1 (\theta [\rm{deg}])^2$ meV 
is the effective Landau level splitting
and $\theta$ is the twist angle.  The numerical value here is estimated for MoTe$_2$,
but similar values will hold in WSe$_2$.  At typical twist angles, $\theta\sim 3^{\circ}-5^{\circ}$, the Landau level splitting 
is large enough with respect to Landau level mixing to justify projection of the interacting electron Hamiltonian onto the lowest effective Landau level (see Fig. \ref{fig:Effective_Bandstructure}(c) below). 

{\em Lowest Landau level projection}---
Given the periodic effective fields, it is convenient to examine the lowest Landau level (LLL) projection of 
Eq. \eqref{Hamiltonian_alphamodes} in a representation of Landau gauge guiding center 
states $\ket{X}$. The Hamiltonian can be mapped to one for 
LLL holes experiencing a potential \cite{Pfannkuche,Allan_QHE_Hexagonal} with moiré periodicity: 
\begin{align}
    \bra{X'}H\ket{X} = -\frac{\hbar\,\omega_{c}}{2}\,\delta_{X',X} +\sum_{m,{\bm G}_m} \xi_m\,\bra{X'} e^{i\,{\bm G}_m\cdot{\bm r}} \ket{X},
    \label{EffectiveHamiltonian}
\end{align}
where $m$ is a reciprocal lattice vector shell label, ${\bm G}_m$ belongs to shell $m$ and
\begin{align}
   \bra{X'} e^{i\,{\bm G}\cdot{\bm r}} \ket{X} = e^{-|{\bm G}|^2\ell^2/4}e^{\frac{i}{2}G_x(X+X')} \delta_{X',X+G_y\ell^2},
\end{align}
where $\ell$ is the effective magnetic length ($2 \pi \ell^2 B_0=\Phi_0$).
In Eq.~\eqref{EffectiveHamiltonian} the effective periodic potential has contributions from both kinetic and 
potential terms \cite{Supplemental}:
\begin{align}
    \xi_m=&-\frac{\hbar\,e}{2m^*}\,\alpha_+({\bm G}_m)\,G_{m-}-\delta({\bm G}_m)+\Delta({\bm G}_m)\nonumber \\
    &-\frac{e^2}{2m^*}\sum_{\bm G'}\alpha_+({\bm G}_m-{\bm G'})\alpha_-({\bm G'}).
    \label{EffectiveFourier}
\end{align}
Because the $\xi_0$--contribution yields only a constant energy and 
the magnetic form factor $e^{-|{\bm G}|^2\ell^2/4}$ suppresses contributions from higher shells,
the LLL physics is controlled almost entirely by the Fourier coefficient corresponding to the first shell of reciprocal lattice vectors, $\xi_1$.  The LLL electronic structure can be calculated analytically when only $\xi_1$ is non-zero
and yields a band width proportional to $|\xi_1|$ \cite{Supplemental}. We will now demonstrate that magic angle 
behavior occurs when $\xi_1=0$. When this condition is satisfied, the transformed Hamiltonian is equivalent  
to that of interacting holes in an ordinary Landau level and states in the same universality class as the Laughlin state are expected for fillings $1/m$.

\begin{figure}[h!]
\centering
\includegraphics[width=\linewidth]{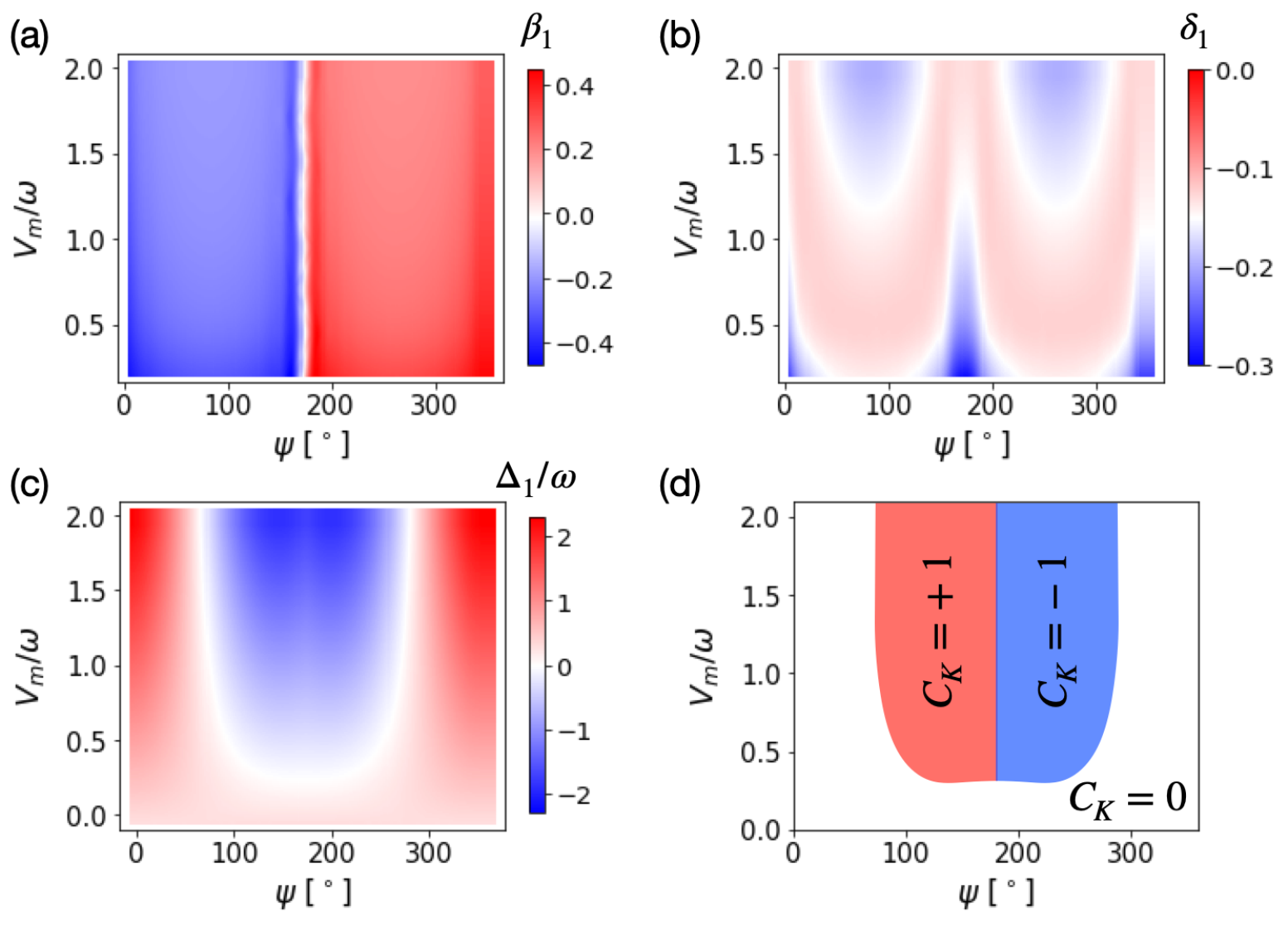}
\caption{Dependence of the first-shell Fourier coefficients (a) $\beta_1$, (b) $\delta_1$ and (c) $\Delta_1$, on the continuum model parameter $\psi$ and $V_m/\omega$, the ratio of the potential and tunneling moir\'e modulation strengths. The units are the same as in Fig. \ref{fig:Effective_MagField}. 
(d) Chern number of the topmost moiré band from the continuum model as a function of $\psi$ and $V_m/\omega$ at $\theta=2.5^{\circ}$. 
The regions with $C_K=\pm 1$ coincide with region where $\Delta_1<0$.}
\label{fig:Effective_Fourier}
\end{figure}

\begin{figure}[h!]
\centering
\includegraphics[width=0.85\linewidth]{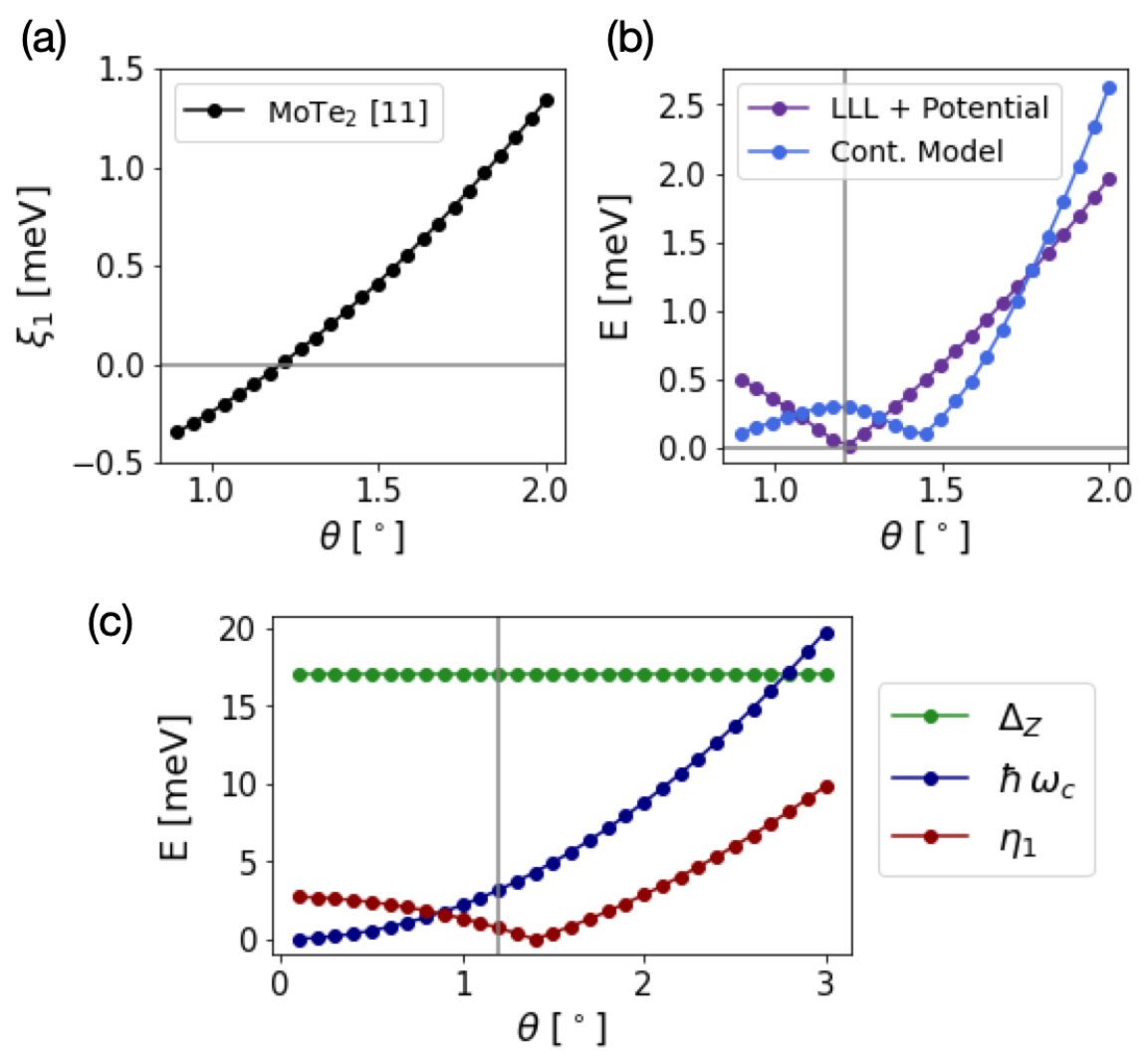}
\caption{(a) First Fourier coefficient of the effective periodic potential $\xi_1$ as a function of twist angle. (b) Band width of the topmost moiré valence band from the continuum model and from our effective LLL in a periodic potential model as a function of twist angle. (c) Comparison, as a function of twist angle, between the effective Zeeman splitting $\Delta_{Z}$, the effective Landau level splitting $\hbar\omega_c$ and the $n=1$ LL--mixing scale $\eta_1=6|\xi^{(1,0)}_1|\sqrt{2\pi}\exp(-\pi/\sqrt{3})/3^{1/4}$ \cite{Supplemental}. The vertical line indicates the magic angle ($\xi_1=0$). These results are for unstrained MoTe$_2$ \cite{FengchengTopology}: $V_m=8$ meV, $\psi=89.6^{\circ}$, $\omega=-8.5$ meV.}
\label{fig:Effective_Bandstructure}
\end{figure}

{\em Magic Angles}---
Using Eq.~\eqref{eq:alphapm} and keeping only the $m=1$ contribution, the coefficient in Eq. \eqref{EffectiveFourier} simplifies to
\begin{align}
    \xi_1 = \hbar\omega_c\left( \frac{|\bar\beta_1|}{2} -\frac{\sqrt{3}\bar\beta_1^2}{8\,\pi}-\bar{\delta}_1\right)+\Delta_1,
    \label{Xi1_Fourier}
\end{align}
where 
$\bar{\beta}_1$ and $\bar{\delta}_1$ are dimensionless \cite{Supplemental}.
Figs. \ref{fig:Effective_Fourier}(a)-(c) show the dependence of $\Delta_1$, $\delta_1$ and $\beta_1$ on the shape parameter $\psi$ and on the ratio $V_m/\omega$.  
The coefficient of $\hbar \omega_c$ in Eq.~\eqref{Xi1_Fourier} is always positive because the pseudospin field direction 
changes most rapidly near the $m$-points in the unit cell \cite{Supplemental}. Since $\hbar \omega_c \propto \theta^2$ and $\Delta_1$ is independent of $\theta$, it follows that $\xi_1$ can cross zero as a function of twist angle only if $\Delta_1$ is negative. Because interlayer tunneling is strong near the $\gamma$ points in the moir\'e cell, it makes a positive 
contribution to $\Delta_1$. In order for $\Delta_1$ to be negative, there must be a large contribution to $\bf{\Delta}$ from the 
moir\'e modulation potential at the $\kappa$-points. These observations explain the dependence of $\Delta_1$ on $V_m/w$ in Fig. \ref{fig:Effective_Fourier}(c), from which we conclude that magic angles will normally appear
for $V_m/w \gtrsim 0.6$. As seen in Fig. \ref{fig:Effective_Fourier}(d), $\Delta_1$ changes sign at approximately the same value of $\psi$ as the Chern number of the topmost moiré band changes from zero to one \cite{FengchengTopology2,Supplemental}, illustrating that the shape of the Skyrmion texture is critical for the formation of topological bands in TMD homobilayers.

In Fig. \ref{fig:Effective_Bandstructure}(a) we plot as an example the evolution of $\xi_1$ with twist angle
for a model \cite{FengchengTopology} of unstrained MoTe$_2$ bilayers. In Fig. \ref{fig:Effective_Bandstructure}(b) the  
band width of the adiabatic approximation effective LLL calculated directly from Eq.~\eqref{Xi1_Fourier} is compared to the corresponding continuum model band width, showing good agreement for the location of the magic angles. When the continuum model is improved by 
accounting for structural relaxation \cite{FCI_DiXiao}, the resulting magic angle is closer to experimental values \cite{FCI_Experiment1,FCI_Experiment2}, $\theta\approx3.75^{\circ}$. The cancellation between Zeeman and kinetic energy terms in the 
Hamiltonian is reminiscent of a similar cancellation that occurs for arbitrary magnetic field distributions
in two-dimensional electron gases when the ratio of the Zeeman spin-splitting to 
$\hbar \omega_c$ equals one, as first observed by Aharonov and Casher \cite{aharonov1979ground}. In the supplemental material \cite{Supplemental} we give an alternative version of the magic angle argument that is related to the Aharonov-Casher cancellation \cite{aharonov1979ground,CrepelChiral}. It implies that our criterion for ideal flat Chern band formation is accurate, even when LL--mixing is not negligible.

Finally, we note that the band width of the LLL effective model goes to a finite value $\propto |\Delta_1|$ in the limit $\theta\to 0$, while for the continuum model the band width vanishes in the same limit, emphasizing that Landau level mixing is essential at very small twist angles.
Fig. \ref{fig:Effective_Bandstructure}(c) shows the effective  LL--splitting $\hbar\omega_c$ and the energy scale of LL-mixing with the $n=1$ LL, $\eta_1$, as a function of twist angle. Fig. \ref{fig:Effective_Bandstructure}(c) also shows a lower bound for the effective Zeeman splitting $\Delta_Z=2\omega$, that provides an estimation of the range of twist angles where the adiabatic approximation holds.

{\em Discussion}---
In this Letter we have presented an analysis of K-valley twisted TMD homobilayers
that is motivated by the presence \cite{FengchengTopology,FengchengTopology2} of 
Skyrmions in the layer pseudospin field of their continuum model Hamiltonians.
In an adiabatic approximation, the Skyrmions give rise to a spatially periodic effective magnetic field 
in the valley projected Hamiltonian 
with one flux quantum per unit cell and a spacing between Landau levels that grows like the square of the twist angle. 
We show by explicit calculation that the magic angle behavior \cite{LiangFuMagic,KaiSun_FCI,Crepel_FCI,FCI_Flatiron} 
thought to be associated with the recently observed FCI \cite{FCI_Experiment1,FCI_Experiment2} states
occurs when the effective periodic potential within the lowest effective Landau level vanishes. The transformation to a Landau level representation explains that 
the trace condition is almost satisfied in the vicinity of the magic angle because the moiré bands inherit the 
ideal quantum geometry of the LLL. It also brings the knowledge gained from decades of studies of the 
conventional fractional quantum Hall effect to bear on the moiré FCI problem.
For example, the fractional charge 
gaps of moiré FCI states in the absence of disorder should be $\sim 0.1 e^2/\epsilon_{hBN} \ell \sim 0.25 e^2/\epsilon_{hBN} \sqrt{A_M}
\sim 10$meV. 

Our approach allows external magnetic fields, which are important for the Streda formula identification \cite{FCI_Experiment1,FCI_Experiment2} of the Chern insulator states,
to be easily incorporated in the theory, simply by adding an external potential contribution to the average field ${\bm B}_0$.
The external field will add to the Landau level degeneracy in one valley and decrease the Landau level degeneracy 
in the other valley, and add a preference for states that are valley polarized in the sense that aligns the orbital 
magnetism with the magnetic field.  At a given effective Landau level filling factor, increasing the effective magnetic field
will increase the interaction energy scale $e^2/\epsilon \ell$, allowing interactions to compete more strongly against 
effective magnetic fields. The Landau level approach to topological moiré TMDs introduced here also simplifies the treatment of the competition between interactions and 
both periodic and random disorder potentials. In general both periodic and random potentials will give the Landau levels a finite energy width, 
which will compete with the electron-electron interactions to determine the ground state at a particular band filling $\nu$. 
This competition likely explains why FCI ground states are measured only at some filling factors.

It is interesting to speculate on what new frontiers in fractional Hall physics might follow from the 
observation of FCI states in K-valley twisted homobilayers.  For instance, it is natural to expect the competition between 
density-wave and incompressible states that is prominent in higher Landau levels \cite{CDW_Fogler,CDW_Moessner,CDW_Haldane,CDW_Yoshioka} to be altered.  Most intriguingly, the effective magnetic field helps decrease magnetic lengths and increase interaction strengths
beyond what is otherwise achievable.  When combined with the possibility of exposing these two-dimensional 
electron systems to scanning probes by eliminating boron nitride encapsulation, this advance brings the prospects for manipulation of 
fractionalized quasiparticles much closer to reality. We leave all these interesting directions for future work.

{\em Acknowledgments ---} We thank Liang Fu and Eslam Khalaf for useful discussions. This work was supported by the U.S. Department of Energy Office of Basic Energy Sciences under Award DE-SC0019481.

\bibliography{refs}

\begin{thebibliography}{44}%
\makeatletter
\providecommand \@ifxundefined [1]{%
 \@ifx{#1\undefined}
}%
\providecommand \@ifnum [1]{%
 \ifnum #1\expandafter \@firstoftwo
 \else \expandafter \@secondoftwo
 \fi
}%
\providecommand \@ifx [1]{%
 \ifx #1\expandafter \@firstoftwo
 \else \expandafter \@secondoftwo
 \fi
}%
\providecommand \natexlab [1]{#1}%
\providecommand \enquote  [1]{``#1''}%
\providecommand \bibnamefont  [1]{#1}%
\providecommand \bibfnamefont [1]{#1}%
\providecommand \citenamefont [1]{#1}%
\providecommand \href@noop [0]{\@secondoftwo}%
\providecommand \href [0]{\begingroup \@sanitize@url \@href}%
\providecommand \@href[1]{\@@startlink{#1}\@@href}%
\providecommand \@@href[1]{\endgroup#1\@@endlink}%
\providecommand \@sanitize@url [0]{\catcode `\\12\catcode `\$12\catcode
  `\&12\catcode `\#12\catcode `\^12\catcode `\_12\catcode `\%12\relax}%
\providecommand \@@startlink[1]{}%
\providecommand \@@endlink[0]{}%
\providecommand \url  [0]{\begingroup\@sanitize@url \@url }%
\providecommand \@url [1]{\endgroup\@href {#1}{\urlprefix }}%
\providecommand \urlprefix  [0]{URL }%
\providecommand \Eprint [0]{\href }%
\providecommand \doibase [0]{https://doi.org/}%
\providecommand \selectlanguage [0]{\@gobble}%
\providecommand \bibinfo  [0]{\@secondoftwo}%
\providecommand \bibfield  [0]{\@secondoftwo}%
\providecommand \translation [1]{[#1]}%
\providecommand \BibitemOpen [0]{}%
\providecommand \bibitemStop [0]{}%
\providecommand \bibitemNoStop [0]{.\EOS\space}%
\providecommand \EOS [0]{\spacefactor3000\relax}%
\providecommand \BibitemShut  [1]{\csname bibitem#1\endcsname}%
\let\auto@bib@innerbib\@empty
\bibitem [{\citenamefont {Cai}\ \emph {et~al.}(2023)\citenamefont {Cai},
  \citenamefont {Anderson}, \citenamefont {Wang}, \citenamefont {Zhang},
  \citenamefont {Liu}, \citenamefont {Holtzmann}, \citenamefont {Zhang},
  \citenamefont {Fan}, \citenamefont {Taniguchi}, \citenamefont {Watanabe},
  \citenamefont {Ran}, \citenamefont {Cao}, \citenamefont {Fu}, \citenamefont
  {Xiao}, \citenamefont {Yao},\ and\ \citenamefont {Xu}}]{FCI_Experiment1}%
  \BibitemOpen
  \bibfield  {author} {\bibinfo {author} {\bibfnamefont {J.}~\bibnamefont
  {Cai}}, \bibinfo {author} {\bibfnamefont {E.}~\bibnamefont {Anderson}},
  \bibinfo {author} {\bibfnamefont {C.}~\bibnamefont {Wang}}, \bibinfo {author}
  {\bibfnamefont {X.}~\bibnamefont {Zhang}}, \bibinfo {author} {\bibfnamefont
  {X.}~\bibnamefont {Liu}}, \bibinfo {author} {\bibfnamefont {W.}~\bibnamefont
  {Holtzmann}}, \bibinfo {author} {\bibfnamefont {Y.}~\bibnamefont {Zhang}},
  \bibinfo {author} {\bibfnamefont {F.}~\bibnamefont {Fan}}, \bibinfo {author}
  {\bibfnamefont {T.}~\bibnamefont {Taniguchi}}, \bibinfo {author}
  {\bibfnamefont {K.}~\bibnamefont {Watanabe}}, \bibinfo {author}
  {\bibfnamefont {Y.}~\bibnamefont {Ran}}, \bibinfo {author} {\bibfnamefont
  {T.}~\bibnamefont {Cao}}, \bibinfo {author} {\bibfnamefont {L.}~\bibnamefont
  {Fu}}, \bibinfo {author} {\bibfnamefont {D.}~\bibnamefont {Xiao}}, \bibinfo
  {author} {\bibfnamefont {W.}~\bibnamefont {Yao}},\ and\ \bibinfo {author}
  {\bibfnamefont {X.}~\bibnamefont {Xu}},\ }\bibfield  {title} {\bibinfo
  {title} {Signatures of fractional quantum anomalous hall states in twisted
  mote2},\ }\href {https://doi.org/10.1038/s41586-023-06289-w} {\bibfield
  {journal} {\bibinfo  {journal} {Nature}\ }\textbf {\bibinfo {volume} {622}},\
  \bibinfo {pages} {63} (\bibinfo {year} {2023})}\BibitemShut {NoStop}%
\bibitem [{\citenamefont {Zeng}\ \emph {et~al.}(2023)\citenamefont {Zeng},
  \citenamefont {Xia}, \citenamefont {Kang}, \citenamefont {Zhu}, \citenamefont
  {Kn{\"u}ppel}, \citenamefont {Vaswani}, \citenamefont {Watanabe},
  \citenamefont {Taniguchi}, \citenamefont {Mak},\ and\ \citenamefont
  {Shan}}]{FCI_Experiment2}%
  \BibitemOpen
  \bibfield  {author} {\bibinfo {author} {\bibfnamefont {Y.}~\bibnamefont
  {Zeng}}, \bibinfo {author} {\bibfnamefont {Z.}~\bibnamefont {Xia}}, \bibinfo
  {author} {\bibfnamefont {K.}~\bibnamefont {Kang}}, \bibinfo {author}
  {\bibfnamefont {J.}~\bibnamefont {Zhu}}, \bibinfo {author} {\bibfnamefont
  {P.}~\bibnamefont {Kn{\"u}ppel}}, \bibinfo {author} {\bibfnamefont
  {C.}~\bibnamefont {Vaswani}}, \bibinfo {author} {\bibfnamefont
  {K.}~\bibnamefont {Watanabe}}, \bibinfo {author} {\bibfnamefont
  {T.}~\bibnamefont {Taniguchi}}, \bibinfo {author} {\bibfnamefont {K.~F.}\
  \bibnamefont {Mak}},\ and\ \bibinfo {author} {\bibfnamefont {J.}~\bibnamefont
  {Shan}},\ }\bibfield  {title} {\bibinfo {title} {Thermodynamic evidence of
  fractional chern insulator in moir{\'e} mote2},\ }\href
  {https://doi.org/10.1038/s41586-023-06452-3} {\bibfield  {journal} {\bibinfo
  {journal} {Nature}\ }\textbf {\bibinfo {volume} {622}},\ \bibinfo {pages}
  {69} (\bibinfo {year} {2023})}\BibitemShut {NoStop}%
\bibitem [{\citenamefont {Park}\ \emph {et~al.}(2023)\citenamefont {Park},
  \citenamefont {Cai}, \citenamefont {Anderson}, \citenamefont {Zhang},
  \citenamefont {Zhu}, \citenamefont {Liu}, \citenamefont {Wang}, \citenamefont
  {Holtzmann}, \citenamefont {Hu}, \citenamefont {Liu}, \citenamefont
  {Taniguchi}, \citenamefont {Watanabe}, \citenamefont {Chu}, \citenamefont
  {Cao}, \citenamefont {Fu}, \citenamefont {Yao}, \citenamefont {Chang},
  \citenamefont {Cobden}, \citenamefont {Xiao},\ and\ \citenamefont
  {Xu}}]{FCI_Transport1}%
  \BibitemOpen
  \bibfield  {author} {\bibinfo {author} {\bibfnamefont {H.}~\bibnamefont
  {Park}}, \bibinfo {author} {\bibfnamefont {J.}~\bibnamefont {Cai}}, \bibinfo
  {author} {\bibfnamefont {E.}~\bibnamefont {Anderson}}, \bibinfo {author}
  {\bibfnamefont {Y.}~\bibnamefont {Zhang}}, \bibinfo {author} {\bibfnamefont
  {J.}~\bibnamefont {Zhu}}, \bibinfo {author} {\bibfnamefont {X.}~\bibnamefont
  {Liu}}, \bibinfo {author} {\bibfnamefont {C.}~\bibnamefont {Wang}}, \bibinfo
  {author} {\bibfnamefont {W.}~\bibnamefont {Holtzmann}}, \bibinfo {author}
  {\bibfnamefont {C.}~\bibnamefont {Hu}}, \bibinfo {author} {\bibfnamefont
  {Z.}~\bibnamefont {Liu}}, \bibinfo {author} {\bibfnamefont {T.}~\bibnamefont
  {Taniguchi}}, \bibinfo {author} {\bibfnamefont {K.}~\bibnamefont {Watanabe}},
  \bibinfo {author} {\bibfnamefont {J.-H.}\ \bibnamefont {Chu}}, \bibinfo
  {author} {\bibfnamefont {T.}~\bibnamefont {Cao}}, \bibinfo {author}
  {\bibfnamefont {L.}~\bibnamefont {Fu}}, \bibinfo {author} {\bibfnamefont
  {W.}~\bibnamefont {Yao}}, \bibinfo {author} {\bibfnamefont {C.-Z.}\
  \bibnamefont {Chang}}, \bibinfo {author} {\bibfnamefont {D.}~\bibnamefont
  {Cobden}}, \bibinfo {author} {\bibfnamefont {D.}~\bibnamefont {Xiao}},\ and\
  \bibinfo {author} {\bibfnamefont {X.}~\bibnamefont {Xu}},\ }\bibfield
  {title} {\bibinfo {title} {Observation of fractionally quantized anomalous
  hall effect},\ }\href {https://doi.org/10.1038/s41586-023-06536-0} {\bibfield
   {journal} {\bibinfo  {journal} {Nature}\ }\textbf {\bibinfo {volume}
  {622}},\ \bibinfo {pages} {74} (\bibinfo {year} {2023})}\BibitemShut
  {NoStop}%
\bibitem [{\citenamefont {Xu}\ \emph {et~al.}(2023)\citenamefont {Xu},
  \citenamefont {Sun}, \citenamefont {Jia}, \citenamefont {Liu}, \citenamefont
  {Xu}, \citenamefont {Li}, \citenamefont {Gu}, \citenamefont {Watanabe},
  \citenamefont {Taniguchi}, \citenamefont {Tong}, \citenamefont {Jia},
  \citenamefont {Shi}, \citenamefont {Jiang}, \citenamefont {Zhang},
  \citenamefont {Liu},\ and\ \citenamefont {Li}}]{FCI_Transport2}%
  \BibitemOpen
  \bibfield  {author} {\bibinfo {author} {\bibfnamefont {F.}~\bibnamefont
  {Xu}}, \bibinfo {author} {\bibfnamefont {Z.}~\bibnamefont {Sun}}, \bibinfo
  {author} {\bibfnamefont {T.}~\bibnamefont {Jia}}, \bibinfo {author}
  {\bibfnamefont {C.}~\bibnamefont {Liu}}, \bibinfo {author} {\bibfnamefont
  {C.}~\bibnamefont {Xu}}, \bibinfo {author} {\bibfnamefont {C.}~\bibnamefont
  {Li}}, \bibinfo {author} {\bibfnamefont {Y.}~\bibnamefont {Gu}}, \bibinfo
  {author} {\bibfnamefont {K.}~\bibnamefont {Watanabe}}, \bibinfo {author}
  {\bibfnamefont {T.}~\bibnamefont {Taniguchi}}, \bibinfo {author}
  {\bibfnamefont {B.}~\bibnamefont {Tong}}, \bibinfo {author} {\bibfnamefont
  {J.}~\bibnamefont {Jia}}, \bibinfo {author} {\bibfnamefont {Z.}~\bibnamefont
  {Shi}}, \bibinfo {author} {\bibfnamefont {S.}~\bibnamefont {Jiang}}, \bibinfo
  {author} {\bibfnamefont {Y.}~\bibnamefont {Zhang}}, \bibinfo {author}
  {\bibfnamefont {X.}~\bibnamefont {Liu}},\ and\ \bibinfo {author}
  {\bibfnamefont {T.}~\bibnamefont {Li}},\ }\bibfield  {title} {\bibinfo
  {title} {Observation of integer and fractional quantum anomalous hall effects
  in twisted bilayer ${\mathrm{mote}}_{2}$},\ }\href
  {https://doi.org/10.1103/PhysRevX.13.031037} {\bibfield  {journal} {\bibinfo
  {journal} {Phys. Rev. X}\ }\textbf {\bibinfo {volume} {13}},\ \bibinfo
  {pages} {031037} (\bibinfo {year} {2023})}\BibitemShut {NoStop}%
\bibitem [{\citenamefont {Lu}\ \emph {et~al.}(2023)\citenamefont {Lu},
  \citenamefont {Han}, \citenamefont {Yao}, \citenamefont {Reddy},
  \citenamefont {Yang}, \citenamefont {Seo}, \citenamefont {Watanabe},
  \citenamefont {Taniguchi}, \citenamefont {Fu},\ and\ \citenamefont
  {Ju}}]{lu2023fractional}%
  \BibitemOpen
  \bibfield  {author} {\bibinfo {author} {\bibfnamefont {Z.}~\bibnamefont
  {Lu}}, \bibinfo {author} {\bibfnamefont {T.}~\bibnamefont {Han}}, \bibinfo
  {author} {\bibfnamefont {Y.}~\bibnamefont {Yao}}, \bibinfo {author}
  {\bibfnamefont {A.~P.}\ \bibnamefont {Reddy}}, \bibinfo {author}
  {\bibfnamefont {J.}~\bibnamefont {Yang}}, \bibinfo {author} {\bibfnamefont
  {J.}~\bibnamefont {Seo}}, \bibinfo {author} {\bibfnamefont {K.}~\bibnamefont
  {Watanabe}}, \bibinfo {author} {\bibfnamefont {T.}~\bibnamefont {Taniguchi}},
  \bibinfo {author} {\bibfnamefont {L.}~\bibnamefont {Fu}},\ and\ \bibinfo
  {author} {\bibfnamefont {L.}~\bibnamefont {Ju}},\ }\bibfield  {title}
  {\bibinfo {title} {Fractional quantum anomalous hall effect in a graphene
  moire superlattice},\ }\href@noop {} {\bibfield  {journal} {\bibinfo
  {journal} {arXiv preprint arXiv:2309.17436}\ } (\bibinfo {year}
  {2023})}\BibitemShut {NoStop}%
\bibitem [{Note1()}]{Note1}%
  \BibitemOpen
  \bibinfo {note} {Recently the term fractional Chern insulator has also been
  used \cite {xie2021fractional} to describe fractional quantum Hall states
  that occur in a moiré band at non-zero magnetic field. If this terminology
  is adopted, the zero-field states of interest here, which combine magnetism
  with fractionalization since they break time-reversal spontaneously, should
  be referred to as fractional quantum anomalous Hall states.}\BibitemShut
  {Stop}%
\bibitem [{\citenamefont {Tang}\ \emph {et~al.}(2011)\citenamefont {Tang},
  \citenamefont {Mei},\ and\ \citenamefont {Wen}}]{Wen_FCI}%
  \BibitemOpen
  \bibfield  {author} {\bibinfo {author} {\bibfnamefont {E.}~\bibnamefont
  {Tang}}, \bibinfo {author} {\bibfnamefont {J.-W.}\ \bibnamefont {Mei}},\ and\
  \bibinfo {author} {\bibfnamefont {X.-G.}\ \bibnamefont {Wen}},\ }\bibfield
  {title} {\bibinfo {title} {High-temperature fractional quantum hall states},\
  }\href {https://doi.org/10.1103/PhysRevLett.106.236802} {\bibfield  {journal}
  {\bibinfo  {journal} {Phys. Rev. Lett.}\ }\textbf {\bibinfo {volume} {106}},\
  \bibinfo {pages} {236802} (\bibinfo {year} {2011})}\BibitemShut {NoStop}%
\bibitem [{\citenamefont {Sun}\ \emph {et~al.}(2011)\citenamefont {Sun},
  \citenamefont {Gu}, \citenamefont {Katsura},\ and\ \citenamefont
  {Das~Sarma}}]{DasSarma_Sun_FCI}%
  \BibitemOpen
  \bibfield  {author} {\bibinfo {author} {\bibfnamefont {K.}~\bibnamefont
  {Sun}}, \bibinfo {author} {\bibfnamefont {Z.}~\bibnamefont {Gu}}, \bibinfo
  {author} {\bibfnamefont {H.}~\bibnamefont {Katsura}},\ and\ \bibinfo {author}
  {\bibfnamefont {S.}~\bibnamefont {Das~Sarma}},\ }\bibfield  {title} {\bibinfo
  {title} {Nearly flatbands with nontrivial topology},\ }\href
  {https://doi.org/10.1103/PhysRevLett.106.236803} {\bibfield  {journal}
  {\bibinfo  {journal} {Phys. Rev. Lett.}\ }\textbf {\bibinfo {volume} {106}},\
  \bibinfo {pages} {236803} (\bibinfo {year} {2011})}\BibitemShut {NoStop}%
\bibitem [{\citenamefont {Neupert}\ \emph {et~al.}(2011)\citenamefont
  {Neupert}, \citenamefont {Santos}, \citenamefont {Chamon},\ and\
  \citenamefont {Mudry}}]{Neupert_FCI}%
  \BibitemOpen
  \bibfield  {author} {\bibinfo {author} {\bibfnamefont {T.}~\bibnamefont
  {Neupert}}, \bibinfo {author} {\bibfnamefont {L.}~\bibnamefont {Santos}},
  \bibinfo {author} {\bibfnamefont {C.}~\bibnamefont {Chamon}},\ and\ \bibinfo
  {author} {\bibfnamefont {C.}~\bibnamefont {Mudry}},\ }\bibfield  {title}
  {\bibinfo {title} {Fractional quantum hall states at zero magnetic field},\
  }\href {https://doi.org/10.1103/PhysRevLett.106.236804} {\bibfield  {journal}
  {\bibinfo  {journal} {Phys. Rev. Lett.}\ }\textbf {\bibinfo {volume} {106}},\
  \bibinfo {pages} {236804} (\bibinfo {year} {2011})}\BibitemShut {NoStop}%
\bibitem [{\citenamefont {Regnault}\ and\ \citenamefont
  {Bernevig}(2011)}]{Bernevig_Regnault_FCI}%
  \BibitemOpen
  \bibfield  {author} {\bibinfo {author} {\bibfnamefont {N.}~\bibnamefont
  {Regnault}}\ and\ \bibinfo {author} {\bibfnamefont {B.~A.}\ \bibnamefont
  {Bernevig}},\ }\bibfield  {title} {\bibinfo {title} {Fractional chern
  insulator},\ }\href {https://doi.org/10.1103/PhysRevX.1.021014} {\bibfield
  {journal} {\bibinfo  {journal} {Phys. Rev. X}\ }\textbf {\bibinfo {volume}
  {1}},\ \bibinfo {pages} {021014} (\bibinfo {year} {2011})}\BibitemShut
  {NoStop}%
\bibitem [{\citenamefont {Sheng}\ \emph {et~al.}(2011)\citenamefont {Sheng},
  \citenamefont {Gu}, \citenamefont {Sun},\ and\ \citenamefont
  {Sheng}}]{Sheng_FCI}%
  \BibitemOpen
  \bibfield  {author} {\bibinfo {author} {\bibfnamefont {D.~N.}\ \bibnamefont
  {Sheng}}, \bibinfo {author} {\bibfnamefont {Z.-C.}\ \bibnamefont {Gu}},
  \bibinfo {author} {\bibfnamefont {K.}~\bibnamefont {Sun}},\ and\ \bibinfo
  {author} {\bibfnamefont {L.}~\bibnamefont {Sheng}},\ }\bibfield  {title}
  {\bibinfo {title} {Fractional quantum hall effect in the absence of landau
  levels},\ }\href {https://doi.org/10.1038/ncomms1380} {\bibfield  {journal}
  {\bibinfo  {journal} {Nature Communications}\ }\textbf {\bibinfo {volume}
  {2}},\ \bibinfo {pages} {389} (\bibinfo {year} {2011})}\BibitemShut {NoStop}%
\bibitem [{\citenamefont {Parameswaran}\ \emph {et~al.}(2012)\citenamefont
  {Parameswaran}, \citenamefont {Roy},\ and\ \citenamefont
  {Sondhi}}]{Parameswaran_Ideal}%
  \BibitemOpen
  \bibfield  {author} {\bibinfo {author} {\bibfnamefont {S.~A.}\ \bibnamefont
  {Parameswaran}}, \bibinfo {author} {\bibfnamefont {R.}~\bibnamefont {Roy}},\
  and\ \bibinfo {author} {\bibfnamefont {S.~L.}\ \bibnamefont {Sondhi}},\
  }\bibfield  {title} {\bibinfo {title} {Fractional chern insulators and the
  ${W}_{\ensuremath{\infty}}$ algebra},\ }\href
  {https://doi.org/10.1103/PhysRevB.85.241308} {\bibfield  {journal} {\bibinfo
  {journal} {Phys. Rev. B}\ }\textbf {\bibinfo {volume} {85}},\ \bibinfo
  {pages} {241308} (\bibinfo {year} {2012})}\BibitemShut {NoStop}%
\bibitem [{\citenamefont {Wu}\ \emph {et~al.}(2019)\citenamefont {Wu},
  \citenamefont {Lovorn}, \citenamefont {Tutuc}, \citenamefont {Martin},\ and\
  \citenamefont {MacDonald}}]{FengchengTopology}%
  \BibitemOpen
  \bibfield  {author} {\bibinfo {author} {\bibfnamefont {F.}~\bibnamefont
  {Wu}}, \bibinfo {author} {\bibfnamefont {T.}~\bibnamefont {Lovorn}}, \bibinfo
  {author} {\bibfnamefont {E.}~\bibnamefont {Tutuc}}, \bibinfo {author}
  {\bibfnamefont {I.}~\bibnamefont {Martin}},\ and\ \bibinfo {author}
  {\bibfnamefont {A.~H.}\ \bibnamefont {MacDonald}},\ }\bibfield  {title}
  {\bibinfo {title} {Topological insulators in twisted transition metal
  dichalcogenide homobilayers},\ }\href
  {https://doi.org/10.1103/PhysRevLett.122.086402} {\bibfield  {journal}
  {\bibinfo  {journal} {Phys. Rev. Lett.}\ }\textbf {\bibinfo {volume} {122}},\
  \bibinfo {pages} {086402} (\bibinfo {year} {2019})}\BibitemShut {NoStop}%
\bibitem [{\citenamefont {Pan}\ \emph {et~al.}(2020)\citenamefont {Pan},
  \citenamefont {Wu},\ and\ \citenamefont {Das~Sarma}}]{FengchengTopology2}%
  \BibitemOpen
  \bibfield  {author} {\bibinfo {author} {\bibfnamefont {H.}~\bibnamefont
  {Pan}}, \bibinfo {author} {\bibfnamefont {F.}~\bibnamefont {Wu}},\ and\
  \bibinfo {author} {\bibfnamefont {S.}~\bibnamefont {Das~Sarma}},\ }\bibfield
  {title} {\bibinfo {title} {Band topology, hubbard model, heisenberg model,
  and dzyaloshinskii-moriya interaction in twisted bilayer
  ${\mathrm{wse}}_{2}$},\ }\href
  {https://doi.org/10.1103/PhysRevResearch.2.033087} {\bibfield  {journal}
  {\bibinfo  {journal} {Phys. Rev. Res.}\ }\textbf {\bibinfo {volume} {2}},\
  \bibinfo {pages} {033087} (\bibinfo {year} {2020})}\BibitemShut {NoStop}%
\bibitem [{\citenamefont {Devakul}\ \emph {et~al.}(2021)\citenamefont
  {Devakul}, \citenamefont {Cr{\'e}pel}, \citenamefont {Zhang},\ and\
  \citenamefont {Fu}}]{LiangFuMagic}%
  \BibitemOpen
  \bibfield  {author} {\bibinfo {author} {\bibfnamefont {T.}~\bibnamefont
  {Devakul}}, \bibinfo {author} {\bibfnamefont {V.}~\bibnamefont {Cr{\'e}pel}},
  \bibinfo {author} {\bibfnamefont {Y.}~\bibnamefont {Zhang}},\ and\ \bibinfo
  {author} {\bibfnamefont {L.}~\bibnamefont {Fu}},\ }\bibfield  {title}
  {\bibinfo {title} {Magic in twisted transition metal dichalcogenide
  bilayers},\ }\href {https://doi.org/10.1038/s41467-021-27042-9} {\bibfield
  {journal} {\bibinfo  {journal} {Nature Communications}\ }\textbf {\bibinfo
  {volume} {12}},\ \bibinfo {pages} {6730} (\bibinfo {year}
  {2021})}\BibitemShut {NoStop}%
\bibitem [{\citenamefont {Li}\ \emph {et~al.}(2021)\citenamefont {Li},
  \citenamefont {Kumar}, \citenamefont {Sun},\ and\ \citenamefont
  {Lin}}]{KaiSun_FCI}%
  \BibitemOpen
  \bibfield  {author} {\bibinfo {author} {\bibfnamefont {H.}~\bibnamefont
  {Li}}, \bibinfo {author} {\bibfnamefont {U.}~\bibnamefont {Kumar}}, \bibinfo
  {author} {\bibfnamefont {K.}~\bibnamefont {Sun}},\ and\ \bibinfo {author}
  {\bibfnamefont {S.-Z.}\ \bibnamefont {Lin}},\ }\bibfield  {title} {\bibinfo
  {title} {Spontaneous fractional chern insulators in transition metal
  dichalcogenide moir\'e superlattices},\ }\href
  {https://doi.org/10.1103/PhysRevResearch.3.L032070} {\bibfield  {journal}
  {\bibinfo  {journal} {Phys. Rev. Research}\ }\textbf {\bibinfo {volume}
  {3}},\ \bibinfo {pages} {L032070} (\bibinfo {year} {2021})}\BibitemShut
  {NoStop}%
\bibitem [{\citenamefont {Cr\'epel}\ and\ \citenamefont
  {Fu}(2023)}]{Crepel_FCI}%
  \BibitemOpen
  \bibfield  {author} {\bibinfo {author} {\bibfnamefont {V.}~\bibnamefont
  {Cr\'epel}}\ and\ \bibinfo {author} {\bibfnamefont {L.}~\bibnamefont {Fu}},\
  }\bibfield  {title} {\bibinfo {title} {Anomalous hall metal and fractional
  chern insulator in twisted transition metal dichalcogenides},\ }\href
  {https://doi.org/10.1103/PhysRevB.107.L201109} {\bibfield  {journal}
  {\bibinfo  {journal} {Phys. Rev. B}\ }\textbf {\bibinfo {volume} {107}},\
  \bibinfo {pages} {L201109} (\bibinfo {year} {2023})}\BibitemShut {NoStop}%
\bibitem [{\citenamefont {Morales-Dur\'an}\ \emph {et~al.}(2023)\citenamefont
  {Morales-Dur\'an}, \citenamefont {Wang}, \citenamefont {Schleder},
  \citenamefont {Angeli}, \citenamefont {Zhu}, \citenamefont {Kaxiras},
  \citenamefont {Repellin},\ and\ \citenamefont {Cano}}]{FCI_Flatiron}%
  \BibitemOpen
  \bibfield  {author} {\bibinfo {author} {\bibfnamefont {N.}~\bibnamefont
  {Morales-Dur\'an}}, \bibinfo {author} {\bibfnamefont {J.}~\bibnamefont
  {Wang}}, \bibinfo {author} {\bibfnamefont {G.~R.}\ \bibnamefont {Schleder}},
  \bibinfo {author} {\bibfnamefont {M.}~\bibnamefont {Angeli}}, \bibinfo
  {author} {\bibfnamefont {Z.}~\bibnamefont {Zhu}}, \bibinfo {author}
  {\bibfnamefont {E.}~\bibnamefont {Kaxiras}}, \bibinfo {author} {\bibfnamefont
  {C.}~\bibnamefont {Repellin}},\ and\ \bibinfo {author} {\bibfnamefont
  {J.}~\bibnamefont {Cano}},\ }\bibfield  {title} {\bibinfo {title}
  {Pressure-enhanced fractional chern insulators along a magic line in moir\'e
  transition metal dichalcogenides},\ }\href
  {https://doi.org/10.1103/PhysRevResearch.5.L032022} {\bibfield  {journal}
  {\bibinfo  {journal} {Phys. Rev. Res.}\ }\textbf {\bibinfo {volume} {5}},\
  \bibinfo {pages} {L032022} (\bibinfo {year} {2023})}\BibitemShut {NoStop}%
\bibitem [{Note2()}]{Note2}%
  \BibitemOpen
  \bibinfo {note} {We do not specifically address the FQAH very recently
  observed \cite {lu2023fractional} in rhombohedral graphene
  stacks.}\BibitemShut {Stop}%
\bibitem [{\citenamefont {Wu}\ \emph {et~al.}(2018)\citenamefont {Wu},
  \citenamefont {Lovorn}, \citenamefont {Tutuc},\ and\ \citenamefont
  {MacDonald}}]{wu2018hubbard}%
  \BibitemOpen
  \bibfield  {author} {\bibinfo {author} {\bibfnamefont {F.}~\bibnamefont
  {Wu}}, \bibinfo {author} {\bibfnamefont {T.}~\bibnamefont {Lovorn}}, \bibinfo
  {author} {\bibfnamefont {E.}~\bibnamefont {Tutuc}},\ and\ \bibinfo {author}
  {\bibfnamefont {A.~H.}\ \bibnamefont {MacDonald}},\ }\bibfield  {title}
  {\bibinfo {title} {Hubbard model physics in transition metal dichalcogenide
  moir{\'e} bands},\ }\href@noop {} {\bibfield  {journal} {\bibinfo  {journal}
  {Physical review letters}\ }\textbf {\bibinfo {volume} {121}},\ \bibinfo
  {pages} {026402} (\bibinfo {year} {2018})}\BibitemShut {NoStop}%
\bibitem [{\citenamefont {Wang}\ \emph {et~al.}(2023)\citenamefont {Wang},
  \citenamefont {Zhang}, \citenamefont {Liu}, \citenamefont {He}, \citenamefont
  {Xu}, \citenamefont {Ran}, \citenamefont {Cao},\ and\ \citenamefont
  {Xiao}}]{FCI_DiXiao}%
  \BibitemOpen
  \bibfield  {author} {\bibinfo {author} {\bibfnamefont {C.}~\bibnamefont
  {Wang}}, \bibinfo {author} {\bibfnamefont {X.-W.}\ \bibnamefont {Zhang}},
  \bibinfo {author} {\bibfnamefont {X.}~\bibnamefont {Liu}}, \bibinfo {author}
  {\bibfnamefont {Y.}~\bibnamefont {He}}, \bibinfo {author} {\bibfnamefont
  {X.}~\bibnamefont {Xu}}, \bibinfo {author} {\bibfnamefont {Y.}~\bibnamefont
  {Ran}}, \bibinfo {author} {\bibfnamefont {T.}~\bibnamefont {Cao}},\ and\
  \bibinfo {author} {\bibfnamefont {D.}~\bibnamefont {Xiao}},\ }\href@noop {}
  {\bibinfo {title} {Fractional chern insulator in twisted bilayer mote$_2$}}
  (\bibinfo {year} {2023}),\ \Eprint {https://arxiv.org/abs/2304.11864}
  {arXiv:2304.11864 [cond-mat.str-el]} \BibitemShut {NoStop}%
\bibitem [{\citenamefont {Reddy}\ \emph {et~al.}(2023)\citenamefont {Reddy},
  \citenamefont {Alsallom}, \citenamefont {Zhang}, \citenamefont {Devakul},\
  and\ \citenamefont {Fu}}]{FCI_LiangFu}%
  \BibitemOpen
  \bibfield  {author} {\bibinfo {author} {\bibfnamefont {A.~P.}\ \bibnamefont
  {Reddy}}, \bibinfo {author} {\bibfnamefont {F.}~\bibnamefont {Alsallom}},
  \bibinfo {author} {\bibfnamefont {Y.}~\bibnamefont {Zhang}}, \bibinfo
  {author} {\bibfnamefont {T.}~\bibnamefont {Devakul}},\ and\ \bibinfo {author}
  {\bibfnamefont {L.}~\bibnamefont {Fu}},\ }\bibfield  {title} {\bibinfo
  {title} {Fractional quantum anomalous hall states in twisted bilayer
  ${\mathrm{mote}}_{2}$ and ${\mathrm{wse}}_{2}$},\ }\href
  {https://doi.org/10.1103/PhysRevB.108.085117} {\bibfield  {journal} {\bibinfo
   {journal} {Phys. Rev. B}\ }\textbf {\bibinfo {volume} {108}},\ \bibinfo
  {pages} {085117} (\bibinfo {year} {2023})}\BibitemShut {NoStop}%
\bibitem [{Sup()}]{Supplemental}%
  \BibitemOpen
  \href@noop {} {}\bibinfo {note} {See supplemental material for (a) Details on
  the continuum model for homobilayer moir\'e TMDs, (b) The derivation of the
  adiabatic Hamiltonian, (c) Details on the Landau level representation of the
  adiabatic Hamiltonian (d) The derivation of the expression for the magic
  angle and (e) Calculation of the bandstructure for the effective LLL model in
  a periodic potential.}\BibitemShut {Stop}%
\bibitem [{\citenamefont {Ye}\ \emph {et~al.}(1999)\citenamefont {Ye},
  \citenamefont {Kim}, \citenamefont {Millis}, \citenamefont {Shraiman},
  \citenamefont {Majumdar},\ and\ \citenamefont {Te\ifmmode \check{s}\else
  \v{s}\fi{}anovi\ifmmode~\acute{c}\else \'{c}\fi{}}}]{Millis_TopologicalHall}%
  \BibitemOpen
  \bibfield  {author} {\bibinfo {author} {\bibfnamefont {J.}~\bibnamefont
  {Ye}}, \bibinfo {author} {\bibfnamefont {Y.~B.}\ \bibnamefont {Kim}},
  \bibinfo {author} {\bibfnamefont {A.~J.}\ \bibnamefont {Millis}}, \bibinfo
  {author} {\bibfnamefont {B.~I.}\ \bibnamefont {Shraiman}}, \bibinfo {author}
  {\bibfnamefont {P.}~\bibnamefont {Majumdar}},\ and\ \bibinfo {author}
  {\bibfnamefont {Z.}~\bibnamefont {Te\ifmmode \check{s}\else
  \v{s}\fi{}anovi\ifmmode~\acute{c}\else \'{c}\fi{}}},\ }\bibfield  {title}
  {\bibinfo {title} {Berry phase theory of the anomalous hall effect:
  Application to colossal magnetoresistance manganites},\ }\href
  {https://doi.org/10.1103/PhysRevLett.83.3737} {\bibfield  {journal} {\bibinfo
   {journal} {Phys. Rev. Lett.}\ }\textbf {\bibinfo {volume} {83}},\ \bibinfo
  {pages} {3737} (\bibinfo {year} {1999})}\BibitemShut {NoStop}%
\bibitem [{\citenamefont {Ohgushi}\ \emph {et~al.}(2000)\citenamefont
  {Ohgushi}, \citenamefont {Murakami},\ and\ \citenamefont
  {Nagaosa}}]{Nagaosa_TopologicalHall_1}%
  \BibitemOpen
  \bibfield  {author} {\bibinfo {author} {\bibfnamefont {K.}~\bibnamefont
  {Ohgushi}}, \bibinfo {author} {\bibfnamefont {S.}~\bibnamefont {Murakami}},\
  and\ \bibinfo {author} {\bibfnamefont {N.}~\bibnamefont {Nagaosa}},\
  }\bibfield  {title} {\bibinfo {title} {Spin anisotropy and quantum hall
  effect in the kagom\'e lattice: Chiral spin state based on a ferromagnet},\
  }\href {https://doi.org/10.1103/PhysRevB.62.R6065} {\bibfield  {journal}
  {\bibinfo  {journal} {Phys. Rev. B}\ }\textbf {\bibinfo {volume} {62}},\
  \bibinfo {pages} {R6065} (\bibinfo {year} {2000})}\BibitemShut {NoStop}%
\bibitem [{\citenamefont {Hamamoto}\ \emph {et~al.}(2015)\citenamefont
  {Hamamoto}, \citenamefont {Ezawa},\ and\ \citenamefont
  {Nagaosa}}]{Nagaosa_TopologicalHall_2}%
  \BibitemOpen
  \bibfield  {author} {\bibinfo {author} {\bibfnamefont {K.}~\bibnamefont
  {Hamamoto}}, \bibinfo {author} {\bibfnamefont {M.}~\bibnamefont {Ezawa}},\
  and\ \bibinfo {author} {\bibfnamefont {N.}~\bibnamefont {Nagaosa}},\
  }\bibfield  {title} {\bibinfo {title} {Quantized topological hall effect in
  skyrmion crystal},\ }\href {https://doi.org/10.1103/PhysRevB.92.115417}
  {\bibfield  {journal} {\bibinfo  {journal} {Phys. Rev. B}\ }\textbf {\bibinfo
  {volume} {92}},\ \bibinfo {pages} {115417} (\bibinfo {year}
  {2015})}\BibitemShut {NoStop}%
\bibitem [{\citenamefont {van Hoogdalem}\ \emph {et~al.}(2013)\citenamefont
  {van Hoogdalem}, \citenamefont {Tserkovnyak},\ and\ \citenamefont
  {Loss}}]{van2013magnetic}%
  \BibitemOpen
  \bibfield  {author} {\bibinfo {author} {\bibfnamefont {K.~A.}\ \bibnamefont
  {van Hoogdalem}}, \bibinfo {author} {\bibfnamefont {Y.}~\bibnamefont
  {Tserkovnyak}},\ and\ \bibinfo {author} {\bibfnamefont {D.}~\bibnamefont
  {Loss}},\ }\bibfield  {title} {\bibinfo {title} {Magnetic texture-induced
  thermal hall effects},\ }\href@noop {} {\bibfield  {journal} {\bibinfo
  {journal} {Physical Review B}\ }\textbf {\bibinfo {volume} {87}},\ \bibinfo
  {pages} {024402} (\bibinfo {year} {2013})}\BibitemShut {NoStop}%
\bibitem [{\citenamefont {Volovik}(1987)}]{volovik1987linear}%
  \BibitemOpen
  \bibfield  {author} {\bibinfo {author} {\bibfnamefont {G.}~\bibnamefont
  {Volovik}},\ }\bibfield  {title} {\bibinfo {title} {Linear momentum in
  ferromagnets},\ }\href@noop {} {\bibfield  {journal} {\bibinfo  {journal}
  {Journal of Physics C: Solid State Physics}\ }\textbf {\bibinfo {volume}
  {20}},\ \bibinfo {pages} {L83} (\bibinfo {year} {1987})}\BibitemShut
  {NoStop}%
\bibitem [{\citenamefont {Bruno}\ \emph {et~al.}(2004)\citenamefont {Bruno},
  \citenamefont {Dugaev},\ and\ \citenamefont
  {Taillefumier}}]{bruno2004topological}%
  \BibitemOpen
  \bibfield  {author} {\bibinfo {author} {\bibfnamefont {P.}~\bibnamefont
  {Bruno}}, \bibinfo {author} {\bibfnamefont {V.}~\bibnamefont {Dugaev}},\ and\
  \bibinfo {author} {\bibfnamefont {M.}~\bibnamefont {Taillefumier}},\
  }\bibfield  {title} {\bibinfo {title} {Topological hall effect and berry
  phase in magnetic nanostructures},\ }\href@noop {} {\bibfield  {journal}
  {\bibinfo  {journal} {Physical review letters}\ }\textbf {\bibinfo {volume}
  {93}},\ \bibinfo {pages} {096806} (\bibinfo {year} {2004})}\BibitemShut
  {NoStop}%
\bibitem [{\citenamefont {Paul}\ \emph {et~al.}(2023)\citenamefont {Paul},
  \citenamefont {Zhang},\ and\ \citenamefont {Fu}}]{Skyrmions_LiangFu}%
  \BibitemOpen
  \bibfield  {author} {\bibinfo {author} {\bibfnamefont {N.}~\bibnamefont
  {Paul}}, \bibinfo {author} {\bibfnamefont {Y.}~\bibnamefont {Zhang}},\ and\
  \bibinfo {author} {\bibfnamefont {L.}~\bibnamefont {Fu}},\ }\bibfield
  {title} {\bibinfo {title} {Giant proximity exchange and flat chern band in 2d
  magnet-semiconductor heterostructures},\ }\href
  {https://doi.org/10.1126/sciadv.abn1401} {\bibfield  {journal} {\bibinfo
  {journal} {Science Advances}\ }\textbf {\bibinfo {volume} {9}},\ \bibinfo
  {pages} {eabn1401} (\bibinfo {year} {2023})},\ \Eprint
  {https://arxiv.org/abs/https://www.science.org/doi/pdf/10.1126/sciadv.abn1401}
  {https://www.science.org/doi/pdf/10.1126/sciadv.abn1401} \BibitemShut
  {NoStop}%
\bibitem [{\citenamefont {Yu}\ \emph {et~al.}(2019)\citenamefont {Yu},
  \citenamefont {Chen},\ and\ \citenamefont {Yao}}]{Skyrmions_TMD1}%
  \BibitemOpen
  \bibfield  {author} {\bibinfo {author} {\bibfnamefont {H.}~\bibnamefont
  {Yu}}, \bibinfo {author} {\bibfnamefont {M.}~\bibnamefont {Chen}},\ and\
  \bibinfo {author} {\bibfnamefont {W.}~\bibnamefont {Yao}},\ }\bibfield
  {title} {\bibinfo {title} {{Giant magnetic field from moiré induced Berry
  phase in homobilayer semiconductors}},\ }\href
  {https://doi.org/10.1093/nsr/nwz117} {\bibfield  {journal} {\bibinfo
  {journal} {National Science Review}\ }\textbf {\bibinfo {volume} {7}},\
  \bibinfo {pages} {12} (\bibinfo {year} {2019})},\ \Eprint
  {https://arxiv.org/abs/https://academic.oup.com/nsr/article-pdf/7/1/12/40810220/nsr\_7\_1\_12.pdf}
  {https://academic.oup.com/nsr/article-pdf/7/1/12/40810220/nsr\_7\_1\_12.pdf}
  \BibitemShut {NoStop}%
\bibitem [{\citenamefont {Zhai}\ and\ \citenamefont
  {Yao}(2020)}]{Skyrmions_TMD2}%
  \BibitemOpen
  \bibfield  {author} {\bibinfo {author} {\bibfnamefont {D.}~\bibnamefont
  {Zhai}}\ and\ \bibinfo {author} {\bibfnamefont {W.}~\bibnamefont {Yao}},\
  }\bibfield  {title} {\bibinfo {title} {Theory of tunable flux lattices in the
  homobilayer moir\'e of twisted and uniformly strained transition metal
  dichalcogenides},\ }\href {https://doi.org/10.1103/PhysRevMaterials.4.094002}
  {\bibfield  {journal} {\bibinfo  {journal} {Phys. Rev. Mater.}\ }\textbf
  {\bibinfo {volume} {4}},\ \bibinfo {pages} {094002} (\bibinfo {year}
  {2020})}\BibitemShut {NoStop}%
\bibitem [{\citenamefont {Pfannkuche}\ and\ \citenamefont
  {Gerhardts}(1992)}]{Pfannkuche}%
  \BibitemOpen
  \bibfield  {author} {\bibinfo {author} {\bibfnamefont {D.}~\bibnamefont
  {Pfannkuche}}\ and\ \bibinfo {author} {\bibfnamefont {R.~R.}\ \bibnamefont
  {Gerhardts}},\ }\bibfield  {title} {\bibinfo {title} {Theory of
  magnetotransport in two-dimensional electron systems subjected to weak
  two-dimensional superlattice potentials},\ }\href
  {https://doi.org/10.1103/PhysRevB.46.12606} {\bibfield  {journal} {\bibinfo
  {journal} {Phys. Rev. B}\ }\textbf {\bibinfo {volume} {46}},\ \bibinfo
  {pages} {12606} (\bibinfo {year} {1992})}\BibitemShut {NoStop}%
\bibitem [{\citenamefont {MacDonald}(1984)}]{Allan_QHE_Hexagonal}%
  \BibitemOpen
  \bibfield  {author} {\bibinfo {author} {\bibfnamefont {A.~H.}\ \bibnamefont
  {MacDonald}},\ }\bibfield  {title} {\bibinfo {title} {Quantized hall effect
  in a hexagonal periodic potential},\ }\href
  {https://doi.org/10.1103/PhysRevB.29.3057} {\bibfield  {journal} {\bibinfo
  {journal} {Phys. Rev. B}\ }\textbf {\bibinfo {volume} {29}},\ \bibinfo
  {pages} {3057} (\bibinfo {year} {1984})}\BibitemShut {NoStop}%
\bibitem [{\citenamefont {Aharonov}\ and\ \citenamefont
  {Casher}(1979)}]{aharonov1979ground}%
  \BibitemOpen
  \bibfield  {author} {\bibinfo {author} {\bibfnamefont {Y.}~\bibnamefont
  {Aharonov}}\ and\ \bibinfo {author} {\bibfnamefont {A.}~\bibnamefont
  {Casher}},\ }\bibfield  {title} {\bibinfo {title} {Ground state of a
  spin-$1/2$ charged particle in a two-dimensional magnetic field},\
  }\href@noop {} {\bibfield  {journal} {\bibinfo  {journal} {Physical Review
  A}\ }\textbf {\bibinfo {volume} {19}},\ \bibinfo {pages} {2461} (\bibinfo
  {year} {1979})}\BibitemShut {NoStop}%
\bibitem [{\citenamefont {Crépel}\ \emph {et~al.}(2023)\citenamefont
  {Crépel}, \citenamefont {Regnault},\ and\ \citenamefont
  {Queiroz}}]{CrepelChiral}%
  \BibitemOpen
  \bibfield  {author} {\bibinfo {author} {\bibfnamefont {V.}~\bibnamefont
  {Crépel}}, \bibinfo {author} {\bibfnamefont {N.}~\bibnamefont {Regnault}},\
  and\ \bibinfo {author} {\bibfnamefont {R.}~\bibnamefont {Queiroz}},\
  }\href@noop {} {\bibinfo {title} {The chiral limits of moir\'e
  semiconductors: origin of flat bands and topology in twisted transition metal
  dichalcogenides homobilayers}} (\bibinfo {year} {2023}),\ \Eprint
  {https://arxiv.org/abs/2305.10477} {arXiv:2305.10477 [cond-mat.mes-hall]}
  \BibitemShut {NoStop}%
\bibitem [{\citenamefont {Koulakov}\ \emph {et~al.}(1996)\citenamefont
  {Koulakov}, \citenamefont {Fogler},\ and\ \citenamefont
  {Shklovskii}}]{CDW_Fogler}%
  \BibitemOpen
  \bibfield  {author} {\bibinfo {author} {\bibfnamefont {A.~A.}\ \bibnamefont
  {Koulakov}}, \bibinfo {author} {\bibfnamefont {M.~M.}\ \bibnamefont
  {Fogler}},\ and\ \bibinfo {author} {\bibfnamefont {B.~I.}\ \bibnamefont
  {Shklovskii}},\ }\bibfield  {title} {\bibinfo {title} {Charge density wave in
  two-dimensional electron liquid in weak magnetic field},\ }\href
  {https://doi.org/10.1103/PhysRevLett.76.499} {\bibfield  {journal} {\bibinfo
  {journal} {Phys. Rev. Lett.}\ }\textbf {\bibinfo {volume} {76}},\ \bibinfo
  {pages} {499} (\bibinfo {year} {1996})}\BibitemShut {NoStop}%
\bibitem [{\citenamefont {Moessner}\ and\ \citenamefont
  {Chalker}(1996)}]{CDW_Moessner}%
  \BibitemOpen
  \bibfield  {author} {\bibinfo {author} {\bibfnamefont {R.}~\bibnamefont
  {Moessner}}\ and\ \bibinfo {author} {\bibfnamefont {J.~T.}\ \bibnamefont
  {Chalker}},\ }\bibfield  {title} {\bibinfo {title} {Exact results for
  interacting electrons in high landau levels},\ }\href
  {https://doi.org/10.1103/PhysRevB.54.5006} {\bibfield  {journal} {\bibinfo
  {journal} {Phys. Rev. B}\ }\textbf {\bibinfo {volume} {54}},\ \bibinfo
  {pages} {5006} (\bibinfo {year} {1996})}\BibitemShut {NoStop}%
\bibitem [{\citenamefont {Haldane}\ \emph {et~al.}(2000)\citenamefont
  {Haldane}, \citenamefont {Rezayi},\ and\ \citenamefont {Yang}}]{CDW_Haldane}%
  \BibitemOpen
  \bibfield  {author} {\bibinfo {author} {\bibfnamefont {F.~D.~M.}\
  \bibnamefont {Haldane}}, \bibinfo {author} {\bibfnamefont {E.~H.}\
  \bibnamefont {Rezayi}},\ and\ \bibinfo {author} {\bibfnamefont
  {K.}~\bibnamefont {Yang}},\ }\bibfield  {title} {\bibinfo {title}
  {Spontaneous breakdown of translational symmetry in quantum hall systems:
  Crystalline order in high landau levels},\ }\href
  {https://doi.org/10.1103/PhysRevLett.85.5396} {\bibfield  {journal} {\bibinfo
   {journal} {Phys. Rev. Lett.}\ }\textbf {\bibinfo {volume} {85}},\ \bibinfo
  {pages} {5396} (\bibinfo {year} {2000})}\BibitemShut {NoStop}%
\bibitem [{\citenamefont {Shibata}\ and\ \citenamefont
  {Yoshioka}(2001)}]{CDW_Yoshioka}%
  \BibitemOpen
  \bibfield  {author} {\bibinfo {author} {\bibfnamefont {N.}~\bibnamefont
  {Shibata}}\ and\ \bibinfo {author} {\bibfnamefont {D.}~\bibnamefont
  {Yoshioka}},\ }\bibfield  {title} {\bibinfo {title} {Ground-state phase
  diagram of 2d electrons in a high landau level: A density-matrix
  renormalization group study},\ }\href
  {https://doi.org/10.1103/PhysRevLett.86.5755} {\bibfield  {journal} {\bibinfo
   {journal} {Phys. Rev. Lett.}\ }\textbf {\bibinfo {volume} {86}},\ \bibinfo
  {pages} {5755} (\bibinfo {year} {2001})}\BibitemShut {NoStop}%
\bibitem [{\citenamefont {Xie}\ \emph {et~al.}(2021)\citenamefont {Xie},
  \citenamefont {Pierce}, \citenamefont {Park}, \citenamefont {Parker},
  \citenamefont {Khalaf}, \citenamefont {Ledwith}, \citenamefont {Cao},
  \citenamefont {Lee}, \citenamefont {Chen}, \citenamefont {Forrester} \emph
  {et~al.}}]{xie2021fractional}%
  \BibitemOpen
  \bibfield  {author} {\bibinfo {author} {\bibfnamefont {Y.}~\bibnamefont
  {Xie}}, \bibinfo {author} {\bibfnamefont {A.~T.}\ \bibnamefont {Pierce}},
  \bibinfo {author} {\bibfnamefont {J.~M.}\ \bibnamefont {Park}}, \bibinfo
  {author} {\bibfnamefont {D.~E.}\ \bibnamefont {Parker}}, \bibinfo {author}
  {\bibfnamefont {E.}~\bibnamefont {Khalaf}}, \bibinfo {author} {\bibfnamefont
  {P.}~\bibnamefont {Ledwith}}, \bibinfo {author} {\bibfnamefont
  {Y.}~\bibnamefont {Cao}}, \bibinfo {author} {\bibfnamefont {S.~H.}\
  \bibnamefont {Lee}}, \bibinfo {author} {\bibfnamefont {S.}~\bibnamefont
  {Chen}}, \bibinfo {author} {\bibfnamefont {P.~R.}\ \bibnamefont {Forrester}},
  \emph {et~al.},\ }\bibfield  {title} {\bibinfo {title} {Fractional chern
  insulators in magic-angle twisted bilayer graphene},\ }\href@noop {}
  {\bibfield  {journal} {\bibinfo  {journal} {Nature}\ }\textbf {\bibinfo
  {volume} {600}},\ \bibinfo {pages} {439} (\bibinfo {year}
  {2021})}\BibitemShut {NoStop}%
\bibitem [{\citenamefont {Laughlin}(1983)}]{Laughlin}%
  \BibitemOpen
  \bibfield  {author} {\bibinfo {author} {\bibfnamefont {R.~B.}\ \bibnamefont
  {Laughlin}},\ }\bibfield  {title} {\bibinfo {title} {Anomalous quantum hall
  effect: An incompressible quantum fluid with fractionally charged
  excitations},\ }\href {https://doi.org/10.1103/PhysRevLett.50.1395}
  {\bibfield  {journal} {\bibinfo  {journal} {Phys. Rev. Lett.}\ }\textbf
  {\bibinfo {volume} {50}},\ \bibinfo {pages} {1395} (\bibinfo {year}
  {1983})}\BibitemShut {NoStop}%
\bibitem [{\citenamefont {Claro}\ and\ \citenamefont {Wannier}(1979)}]{Claro}%
  \BibitemOpen
  \bibfield  {author} {\bibinfo {author} {\bibfnamefont {F.~H.}\ \bibnamefont
  {Claro}}\ and\ \bibinfo {author} {\bibfnamefont {G.~H.}\ \bibnamefont
  {Wannier}},\ }\bibfield  {title} {\bibinfo {title} {Magnetic subband
  structure of electrons in hexagonal lattices},\ }\href
  {https://doi.org/10.1103/PhysRevB.19.6068} {\bibfield  {journal} {\bibinfo
  {journal} {Phys. Rev. B}\ }\textbf {\bibinfo {volume} {19}},\ \bibinfo
  {pages} {6068} (\bibinfo {year} {1979})}\BibitemShut {NoStop}%
\bibitem [{\citenamefont {Yoshioka}(1983)}]{Yoshioka}%
  \BibitemOpen
  \bibfield  {author} {\bibinfo {author} {\bibfnamefont {D.}~\bibnamefont
  {Yoshioka}},\ }\bibfield  {title} {\bibinfo {title} {Hall conductivity of
  two-dimensional electrons in a periodic potential},\ }\href
  {https://doi.org/10.1103/PhysRevB.27.3637} {\bibfield  {journal} {\bibinfo
  {journal} {Phys. Rev. B}\ }\textbf {\bibinfo {volume} {27}},\ \bibinfo
  {pages} {3637} (\bibinfo {year} {1983})}\BibitemShut {NoStop}%
\end{thebibliography}%

\newpage

\appendix
\onecolumngrid
\section*{Supplemental material for ``Magic Angles and Fractional Chern Insulators in Twisted Homobilayer TMDs"}

\section*{Continuum model for AA-stacked moiré TMDs}
The $K$--valley-projected continuum model for AA-stacked TMD homobilayers was introduced in \cite{FengchengTopology}. It is given, in layer space, by
\begin{align}
    H_{TMD}=\begin{pmatrix}
        -\frac{\hbar^2}{2\,m^*}({\bm k}-{\bm k}_b)^2+\Delta_b({\bm r})&\Delta_T({\bm r})\\
        \Delta_T({\bm r})^{\dagger}&-\frac{\hbar^2}{2\,m^*}({\bm k}-{\bm k}_t)^2+\Delta_t({\bm r})
    \end{pmatrix}.
    \label{Sup:ContinuumModel}
\end{align}
The Hamiltonian corresponding to the $K'$--valley is related to this expression via time-reversal symmetry. In Eq. \eqref{Sup:ContinuumModel}, the top and bottom moiré potentials and the interlayer tunneling term are given by
\begin{align}
    \Delta_{b/t}({\bm r})&=2V_m\sum_{j=1,3,5} \cos\left({\bm G}_j\cdot {\bm r} \pm \psi \right)\\
    \Delta_{T}({\bm r})&=\omega\left(1+e^{i {\bm G}_2\cdot {\bm r}}+e^{i {\bm G}_3\cdot {\bm r}}\right),
\end{align}
where $(V_m,\psi,\omega)$ are the material-dependent model parameters introduced in the main text. Due to the rotation between the layers, there is a momentum shift ${\bm k}_{b/t}=k_{\theta}(-1/2,\pm1/2\sqrt{3})$ for the bottom and top dispersions respectively, ${\bm G}_j=k_{\theta}(\cos(\pi j/3),\sin(\pi j/3))$, $j=1,\dots,6$ are the vectors within the first shell of reciprocal lattice vectors and $k_{\theta}=4\pi/\sqrt{3}a_M$. We apply a gauge transformation to remove the relative momentum shift on the diagonal, after which the tunneling is
\begin{align}
    \Delta_T({\bm r})=\omega(e^{i\, {\bm q}_1\cdot {\bm r}}+e^{i\, {\bm q}_2\cdot {\bm r}}+e^{i\, {\bm q}_3\cdot {\bm r}}),
\end{align}
with ${\bm q}_1=k_{\theta}(0,-1/\sqrt{3})$, ${\bm q}_2={\bm G}_2+{\bm q}_1$ and  ${\bm q}_3={\bm G}_3+{\bm q}_1$. We also subtract $\Delta_0=(\Delta_b+\Delta_t)/2$ from the diagonal, to get
\begin{align}
    H_{\text{TMD}}&=\begin{pmatrix}
        -\frac{\hbar^2}{2\,m^*}{\bm k}^2+\frac{\Delta_b-\Delta_t}{2}&\text{Re}\,\Delta_T({\bm r})+i\,\text{Im}\, \Delta_T({\bm r})\\
        \text{Re}\,\Delta_T({\bm r})-i\,\text{Im}\, \Delta_T({\bm r})& -\frac{\hbar^2}{2\,m^*}{\bm k}^2-\frac{\Delta_b-\Delta_t}{2}
    \end{pmatrix}+\begin{pmatrix}
    \frac{\Delta_b(\bm r)+\Delta_t(\bm r)}{2}&0\\
            0&\frac{\Delta_b(\bm r)+\Delta_t(\bm r)}{2}
        \end{pmatrix} \nonumber \\
        &=-\frac{\hbar^2\,{\bm k}^2}{2m^*}\,\sigma_0+{\bm \Delta}({\bm r})\cdot {\bm \sigma}+\Delta_0({\bm r})\,\sigma_0,
\end{align}
which is Eq. (1) in the main text, with ${\bm \Delta}({\bm r})$ defined therein, ${\bm \sigma}=(\sigma_x,\sigma_y,\sigma_z)$ are the layer--Pauli matrices and $\sigma_0$ is the identity matrix.
\begin{figure}[h!]
\centering
\includegraphics[width=0.9\linewidth]{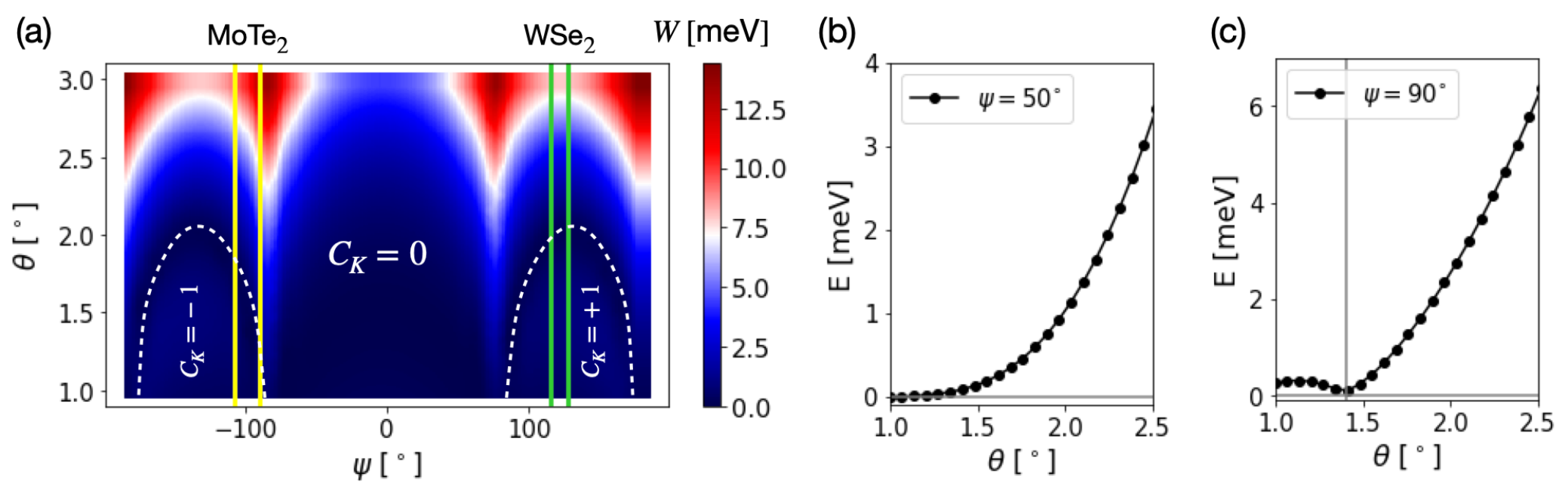}
\caption{(a) Band width of the topmost moiré band obtained from the continuum model Eq. \eqref{Sup:ContinuumModel} as a function of twist angle and shape parameter $\psi$. The two topological regimes are determined by the Chern number of the band vanishing or not. White dashed lines indicate the location of the magic angle within the $C_{K}=\pm 1$ regimes. Green and yellow vertical lines indicate the location of WSe$_2$ and MoTe$_2$ in the phase diagram, according to different references listed in Table I. (b)-(c) Band width as a function of twist angle for $\psi$ in the trivial and topological regimes, respectively. There is a clear qualitative difference, with (c) showing the appearance of a magic angle, indicated by the vertical line. We have used $V_m=8$ meV and $\omega=-8.5$ meV.}
\label{fig:Continuummodel_Phasediagram}
\end{figure}
\begin{center}
    \begin{table}[h]
    \begin{ruledtabular}
    \begin{tabular}{llllll}
    Material&$m^*$($m_0$) & $V_m$ (meV)& $\psi$ ($^{\circ}$)& $\omega$(meV) &Reference \\
    \hline
    MoTe$_2$&0.62&8 &-89.6&-8.5&\cite{FengchengTopology}\\
    MoTe$_2$&0.6&20.8& -107.7 & -23.8& \cite{FCI_DiXiao}\\
    MoTe$_2$&0.62&11.2& -91 & -13.3 & \cite{FCI_LiangFu}\\
    \hline
    WSe$_2$&0.43&9& 128 & -18 & \cite{LiangFuMagic}\\
    WSe$_2$&0.43&6.4& 115.7 & 8.9 & \cite{FCI_Flatiron}
    \end{tabular}
    \end{ruledtabular}
    \caption{Effective mass $m^*$ and continuum model parameters $(V_m,\psi,\omega)$ for MoTe$_2$ and WSe$_2$, obtained from different {\it ab initio} calculations. The lattice constants are $a_0=0.352$ nm for MoTe$_2$ and $a_0=0.332$ nm for WSe$_2$.}
    \end{table}  
\end{center}
The twisted homobilayer TMD continuum model presents two regimes as function of the model parameters $(V_m,\psi,\omega)$, that we refer to as topological and non-topological. In the former case the topmost moiré valence band in valley $K/K'$ has Chern number $C_{K/K'}=\pm 1$, while in the latter the Chern numbers of the topmost moiré band vanish in both valleys $C_{K/K'}=0$. In Fig. \ref{fig:Continuummodel_Phasediagram}(a) we show the band width of the topmost moiré band obtained from diagonalizing the continuum Hamiltonian as a function of twist angle $\theta$ and the continuum model parameter $\psi$, for fixed $V_m=8$ meV and $\omega=-8.5$ meV. In the non-topological regime the band width increases monotonically with $\theta$, as seen in Fig. \ref{fig:Continuummodel_Phasediagram}(b). In contrast, in the topological regime the behavior of band width with twist angle is non-monotonic, giving rise to a {\it magic angle}, Fig. \ref{fig:Continuummodel_Phasediagram}(c), where the band width almost vanishes.\\\\   
As mentioned above, the precise values of continuum model parameters are material-dependent and are obtained from {\it ab initio} calculations. Table I lists continuum model parameters for MoTe$_2$ and WSe$_2$, obtained in different studies. We note that although each DFT calculation predicts different continuum model parameters, all of them place the AA-stacked twisted homobilayers MoTe$_2$ and WSe$_2$ in the topological regime. This can be seen in In Fig. \ref{fig:Continuummodel_Phasediagram}(a), where yellow and green lines indicate the location of MoTe$_2$ \cite{FengchengTopology,FCI_LiangFu,FCI_DiXiao} and WSe$_2$ \cite{LiangFuMagic,FCI_Flatiron} in the topological phase diagram. Because the band width of the resulting topological bands is experimentally tunable, these two moiré materials were proposed as platforms to realize fractional Chern insulators.
\section*{Effective Hamiltonian under the adiabatic approximation}
If we apply the unitary transformation defined in Eq. (2) in the main text to the continuum Hamiltonian for TMD homobilayers, the transformed Hamiltonian is written, in layer space, as
\begin{align}
   H'_{TMD}=U^{\dagger}\,H_{TMD}\,U&=-\frac{1}{2m^*}\begin{pmatrix}
        \hbar {\bm k}+e\tilde{{\bm A}}_{\uparrow\uparrow}&e\tilde{{\bm A}}_{\uparrow\downarrow}\\
        e\tilde{{\bm A}}_{\downarrow\uparrow}&\hbar {\bm k}+e\tilde{{\bm A}}_{\downarrow\downarrow}
    \end{pmatrix}^2+\begin{pmatrix}
        |{\bm \Delta}|+\Delta_0&0\\
        0&-|{\bm \Delta}|+\Delta_0
    \end{pmatrix}.
\end{align}
The non-Abelian connection is $e\,\tilde{\bm A}=-iU^{\dagger}\,\nabla\,U$. Squaring the matrix and noting that $\tilde{{\bm A}}_{\downarrow\uparrow}=\tilde{{\bm A}}_{\uparrow\downarrow}^*$, the transformed Hamiltonian is
\begin{align}
    H'_{TMD}=\begin{pmatrix}
        -\frac{1}{2m^*}(\hbar{\bm k}+e\tilde{{\bm A}}_{\uparrow\uparrow})^2-\frac{e^2}{2m^*}\,|\tilde{{\bm A}}_{\downarrow\uparrow}|^2+\Delta_0+|{\bm \Delta}|&-\frac{e}{2m^*}\tilde{{\bm A}}_{\uparrow\downarrow}(2\hbar{\bm k}+e\tilde{{\bm A}}_{\uparrow\uparrow}+e\tilde{{\bm A}}_{\downarrow\downarrow})\\
        -\frac{e}{2m^*}\tilde{{\bm A}}_{\downarrow\uparrow}(2\hbar{\bm k}+e\tilde{{\bm A}}_{\uparrow\uparrow}+e\tilde{{\bm A}}_{\downarrow\downarrow})&-\frac{1}{2m^*}(\hbar{\bm k}+e\tilde{{\bm A}}_{\downarrow\downarrow})^2-\frac{e^2}{2m^*}\,|\tilde{{\bm A}}_{\downarrow\uparrow}|^2+\Delta_0-|{\bm \Delta}|
    \end{pmatrix}.
\end{align}
In the limit when $|{\bm \Delta}|$ is much larger than all other energy scales, this Hamiltonian corresponds to that of two almost-decoupled pseudospin $\uparrow\uparrow$- and $\downarrow\downarrow$-sectors. Therefore, the low-energy physics of holes can be well approximated by projecting the model to the up pseudospin sector -- the adiabatic approximation -- yielding
\begin{align}
    H'_{TMD}\approx-\frac{1}{2m^*}\left[\hbar{\bm k}+e\tilde{{\bm A}}\right]^2-\frac{e^2}{2m^*}|\tilde{{\bm A}}_{\downarrow\uparrow}|^2+|{\bm \Delta}|+\Delta_0=-\frac{1}{2m^*}\left[\hbar{\bm k}+e\tilde{{\bm A}}\right]^2 - D+\tilde{\Delta}.
    \label{Sup:Adiabatic}
\end{align}
We have used that $|\tilde{{\bm A}}_{\downarrow\uparrow}|^2=(\hbar^2/4e^2)\sum_{i=x,y}\left[\partial_i {\bm n}\right]^2$ and the definition of the kinetic potential introduced in the main text $D=(\hbar^2/8m^*)\sum_{i=x,y}\left[\partial_i {\bm n}\right]^2$. Eq. \eqref{Sup:Adiabatic} is nothing but Eq. (3) in the main text. Note that we omit the layer indices in $\tilde{{\bm A}}$ for shorthand.\\\\
As indicated in the main text, the scalar Hamiltonian Eq. \eqref{Sup:Adiabatic} gains an effective magnetic field that we separate into its homogeneous and periodic parts to obtain Eq. (6) in the main text. After introducing the complex momenta $\Pi_{\pm}=\Pi_x\pm i\Pi_y$ and vector potential $ A_{\pm}=A_x\pm iA_y$, Eq. (6) in the main text can be re-expressed as follows
\begin{align}
    H&=-\frac{\hbar^2}{2m^*}\left[ {\bm \Pi}+\frac{e}{\hbar}\,{\bm{ A}}({\bm r})\right]^2-D({\bm r})+ \tilde{\Delta}({\bm r})\nonumber \\
    &=-\frac{\hbar^2}{2m^*}\left[\frac{1}{2}\left( \Pi_-\Pi_++ \Pi_+ \Pi_-\right)+\frac{e}{\hbar}\left( \Pi_- A_++\Pi_+A_-\right)+\frac{e^2}{\hbar^2}A_- A_+\right]-D+\tilde{\Delta},
\end{align}
where we have used the gauge $\nabla \cdot {\bm A}({\bm r})=0$. We now define the the kinetic-momentum ladder operators for the homogeneous part of the magnetic field (Note the additional $i$-factor),
\begin{align}
 a^{\dagger}=\frac{\ell}{\sqrt{2}}\left( \Pi_y-i\,\Pi_x\right),\qquad  a=\frac{\ell}{\sqrt{2}}\left( \Pi_y+i\,\Pi_x\right),
\end{align}
to cast the Hamiltonian in the form
\begin{align}
    H=-\hbar\omega_c\left( a^{\dagger}a-\frac{1}{2}\right)+\frac{i \,e\,\hbar}{\sqrt{2}\,m^*\,\ell}\left(a\,A_+ - a^{\dagger}\,A_- \right)-\frac{e^2}{2m^*}\,A_-\,A_+-D+\tilde{\Delta}.
\end{align}
Introducing the Fourier expansions of the functions $A_{\pm}(\bm r)$, $D({\bm r})$ and $\tilde{\Delta}(\bm r)$ (Eqs. (7)-(9) in the main text) in the previous equation yields the adiabatic Hamiltonian presented in the main text as Eq. (11). Finally, in order to obtain the Fourier coefficients of the vector potential from those of the effective magnetic field we use $A({\bm k})=i{\bm k}\times {\bm B}({\bm k})/|{\bm k}|^2$. The Fourier expansions of the $x$-- and $y$--components of ${\bm A}$ are given by
\begin{align}
    A_x({\bm r})=\sum_{\bm G}\frac{i\, G_y}{|{\bm G}|^2}\beta({\bm G})e^{i\,{\bm G}\cdot{ \bm r}} \qquad  A_y({\bm r})=\sum_{\bm G}\frac{-i\, G_x}{|{\bm G}|^2}\beta({\bm G})e^{i\,{\bm G}\cdot{ \bm r}},
\end{align}
therefore
\begin{align}
    \alpha_{\pm}({\bm G})=i \frac{G_y}{|{\bm G}|^2}\beta({\bm G})\pm\,i\frac{(-i)\,G_x}{|{\bm G}|^2}\beta({\bm G})=\frac{\pm G_x+i\,G_y}{|{\bm G}|^2}\beta({\bm G}),
    \label{Sup:Acoefficients}
\end{align}
which is Eq. (10) in the main text.
\section*{Adiabatic Hamiltonian in the Landau level basis}
The adiabatic Hamiltonian, Eq. (11) in the main text, is given by
\begin{align}
    H=&-\hbar\omega_c\left( a^{\dagger}a+\frac{1}{2}\right)+\frac{i\,e\,\hbar}{\sqrt{2}\,m^*\ell}\sum_{\bm G}\left(a\,\alpha_+({\bm G})-a^{\dagger}\,\alpha_-({\bm G})\right)\,e^{i\, {\bm G}\cdot {\bm r}}\nonumber \\
    &-\frac{e^2}{2m^*}\, \sum_{{\bm G},{\bm G'}}\alpha_+({\bm G})\,\alpha_-({\bm G'})\,\,e^{i\, ({\bm G}+{\bm G'})\cdot {\bm r}}-\sum_{{\bm G}}\delta({\bm G})\,e^{i\, {\bm G}\cdot {\bm r}}+\sum_{{\bm G}}\Delta({\bm G})\,e^{i\, {\bm G}\cdot {\bm r}}.
    \label{Sup:AdiabaticSupplemental}
\end{align}
The first term of this Hamiltonian has eigenenergies and eigenstates given, in the Landau gauge, by
\begin{align}
    \varepsilon_n=-\hbar \omega_c\left( n+\frac{1}{2}\right) \qquad \braket{{\bm r}|n,X}=\frac{1}{(2^nn!\pi^{1/2}\ell)^{1/2}}\,\text{exp}\left({-\frac{(x-X)^2}{2\ell^2}}\right)H_n\left(\frac{x-X}{\ell}\right),
\end{align}
with $X=k_y\ell^2$ the guiding center and $H_n$ a Hermite polynomial of order $n$. According to \cite{Pfannkuche}, the matrix elements for a plane wave in the Landau gauge basis $\ket{n,X}$ are given by
\begin{align}
    \braket{n',X'|e^{i\, {\bm q}\cdot {\bm r}}|n,X}=e^{-\frac{i}{2}q_x(X+X')}\,\mathcal{L}_{n,n'}({\bm q})\,\delta(X'-X-q_y\ell^2),
\end{align}
where
\begin{align}
    \mathcal{L}_{n,n'}({\bm q})=\left( \frac{m!}{M!}\right)^{\frac{1}{2}}i^{|n'-n|}\left( \frac{q_+}{|{\bm q}|}\right)^{n-n'}\,\left( \frac{|{\bm q}|^2\ell^2}{2}\right)^{|n'-n|/2}\,e^{-q^2\ell^2/4}\,L_m^{(|n'-n|)}\left(\frac{|{\bm q}|^2\ell^2}{2}\right),
\end{align}
with $m=\min(n',n)$, $M=\max(n',n)$ and $L_m^{(\mu)}$ is a generalized Laguerre polynomial. We will use the previous expression to evaluate each term of the adiabatic Hamiltonian Eq. \eqref{Sup:AdiabaticSupplemental}, to cast it in the general form
\begin{align}
        \braket{n',X'|H|n,X}=-\frac{\hbar\,\omega_{c}}{2}\,\delta_{n',n}\,\delta_{X',X} +\sum_{m,{\bm G}_m}\xi^{(n',n)}_m\braket{n',X^{\prime}|e^{i{\bm G}_m\cdot {\bm r}}|n,X}.
\end{align}
Where $m$ labels a reciprocal lattice vector shell and ${\bm G}_m$ belongs to the $m$-th shell of reciprocal lattice vectors.\\\\
{\bf Lowest Landau level projection--}\\
Let us focus first in the projection of the adiabatic Hamiltonian to the LLL, with basis states $\ket{0,X}$. This approximation is justified in the limit of large twist angles, since the Landau level splitting $\hbar\,\omega_c$ is proportional to $\theta^2$. The first crossed term in Eq. \eqref{Sup:AdiabaticSupplemental} is proportional to $ a\,A_+$ and is evaluated as \cite{Pfannkuche,Allan_QHE_Hexagonal}
\begin{align}
    \braket{0,X'|a\sum_{\bm G}\alpha_+({\bm G})\,e^{i\, {\bm G}\cdot {\bm r}}|0,X}&=\sum_{\bm G}\alpha_+({\bm G})\,\braket{1,X'|e^{i\, {\bm G}\cdot {\bm r}}|0,X}\nonumber\\
    &=\sum_{\bm G}\alpha_+({\bm G})\,e^{-|{\bm G}|^2\ell^2/4}e^{-\frac{i}{2}G_x(X+X')}\frac{i\,\ell}{\sqrt{2}}\left(\frac{|G|^2}{G_x+iG_y} \right)\delta\left( X'-X-G_y \ell^2\right).
\end{align}
The second crossed term in Eq. \eqref{Sup:AdiabaticSupplemental} is proportional to $a^{\dagger}\,A_-$ and vanishes due to the action of $a^{\dagger}$ on the LLL state on the left,
\begin{align}
    \braket{0,X'|a^{\dagger}\sum_{\bm G}\alpha_-({\bm G})\,e^{i\, {\bm G}\cdot {\bm r}}|0,X}&=0.
\end{align}
The quadratic term in ${\bm A}$ from Eq. \eqref{Sup:AdiabaticSupplemental} is
\begin{align}
    &\braket{0,X'|\sum_{\bm{G'', G'''}}\alpha_+({\bm G''})\alpha_-({\bm G'''})\,e^{i( \bm{G''+G'''})\cdot{\bm r}}|0,X}\nonumber\\
    &=\sum_{\bm{G'', G'''}}\alpha_+({\bm G''})\alpha_-({\bm G'''})\,e^{-|{\bm G''}+{\bm G'''}|^2\ell^2/4}e^{-\frac{i}{2}(G''_x+G'''_x)(X+X')}\delta\left(X'-X-G''_y\ell^2-G'''_y \ell^2\right)\nonumber\\
    &=\sum_{{\bm G}}\sum_{{\bm G'}}\alpha_+({\bm G}-{\bm G'})\alpha_-({\bm G'})\,e^{-|{\bm G}|^2\ell^2/4}e^{-\frac{i}{2}G_x(X+X')}\delta\left( X'-X-G_y \ell^2\right).
\end{align}
In the last line we defined ${\bm G}={\bm G''}+{\bm G'''}$ and relabelled ${\bm G'''}$ as ${\bm G'}$.\\\\ 
Finally the effective Zeeman field and the kinetic potential $D$ do not involve any LL rising and lowering operators and their projection to the LLL is
\begin{align}
    \braket{0,X'|\sum_{{\bm G}}\Delta({\bm G})\,e^{i\, {\bm G}\cdot {\bm r}}|0,X}=\sum_{{\bm G}}\Delta({\bm G})\,e^{-|{\bm G}|^2\ell^2/4}e^{-\frac{i}{2}G_x(X+X')}\delta\left( X'-X-G_y \ell^2\right).\\
    \braket{0,X'|\sum_{{\bm G}}\delta({\bm G})\,e^{i\, {\bm G}\cdot {\bm r}}|0,X}=\sum_{{\bm G}}\delta({\bm G})\,e^{-|{\bm G}|^2\ell^2/4}e^{-\frac{i}{2}G_x(X+X')}\delta\left( X'-X-G_y \ell^2\right).
\end{align}
Eqs. (34)-(38) can be grouped together to compactly write the adiabatic Hamiltonian projected to the LLL as in Eq. (12) in the main text, with the Fourier coefficients of the effective potential $\xi_m^{(0,0)}$ given by
\begin{align}
    \xi^{(0,0)}_m=&-\frac{\hbar\,e}{2m^*}\,\alpha_+({\bm G}_m)\,G_{m-}
    -\frac{e^2}{2m^*}\sum_{\bm G'}\alpha_+({\bm G}_m-{\bm G'})\alpha_-({\bm G'})-\delta({\bm G}_m)+\Delta({\bm G_m}).
    \label{Sup:EffectiveFourier2}
\end{align}
We note that in the main text, because we are only considering the problem projected to the LLL, we omit the LL-index in the basis states $\ket{0,X}\equiv\ket{X}$ and we adopt the notation $\xi^{(0,0)}_m\equiv \xi_m$.\\\\
{\bf Landau level mixing--}\\
The matrix elements $\braket{n,X^{\prime}|H|0,X}=\braket{0,X^{\prime}|H^{\dagger}|n,X}$ determine the energy scale of LL-mixing between the LLL and the $n$-th LL. These elements can be calculated by evaluating
\begin{align}
    \braket{n,X'|e^{i\, {\bm G}\cdot {\bm r}}|0,X}=e^{-|{\bm G}|^2\ell^2/4}e^{-\frac{i}{2}G_x(X+X')}\delta\left( X'-X-G_y \ell^2\right)\left( \frac{i^n}{\sqrt{n!}}\right)\left(\frac{|G|}{G_+}\right)^n\left(\frac{|G|^2\ell^2}{2}\right)^{n/2},
\end{align}
and noting that
\begin{align}
    \braket{n+1,X'|e^{i\, {\bm G}\cdot {\bm r}}|0,X}=\braket{n,X'|e^{i\, {\bm G}\cdot {\bm r}}|0,X}\left(\frac{i\,G_-\ell}{\sqrt{2}\sqrt{n+1}} \right).
\end{align}
We obtain that the LL-mixing elements are given by
\begin{align}
    \braket{n,X^{\prime}|H|0,X}&=\sum_{m,{\bm G}_m}\xi^{(n,0)}_m\braket{n,X^{\prime}|e^{i{\bm G}_m\cdot {\bm r}}|0,X}, \label{GeneralFourierCoefficents}\\
     \xi^{(n,0)}_m=-\frac{\hbar \,e}{2\,m^*} \left(\frac{1}{n+1}\right)\,\alpha_+({\bm G}_m)G_{m-}&-\frac{\hbar\,e}{m^*\,\ell^2}\,n\,\left(\frac{\alpha_-({\bm G}_m)}{G_{m-}}\right)\nonumber-\frac{e^2}{2m^*}\sum_{\bm G'}\alpha_+({\bm G}_m-{\bm G}^{\prime})\alpha_-({\bm G}^{\prime})-\delta({\bm G}_m)+\Delta({\bm G_m}).
\end{align}
In the following section, when we explicitly evaluate the Fourier coefficients $\xi_m^{(n',n)}$ we will truncate the summation over ${\bm G}'$ to leading order.
\section*{Magic angle conditions}
Because the effective magnetic field has $C_6$-symmetry, all the Fourier coefficients are real and equal within each shell of reciprocal lattice vectors. We denote the Fourier coefficients corresponding to the $i$-th shell in Eqs. (7)-(9) in the main text as $\beta_i$, $\delta_i$ and $\Delta_i$. Plugging Eq. \eqref{Sup:Acoefficients} into Eq. \eqref{Sup:EffectiveFourier2}, we see that the Fourier coefficient from the ${\bm G}=0$ shell gives a constant energy contribution
\begin{align}
    \xi^{(0,0)}_0\equiv\xi_0=-\frac{9\,e^2\,a_M^2}{16\,\pi^2\,m^*}\,\beta_0^2-\delta_0+\Delta_0.
\end{align}
Since the exponential factor $\exp(|{\bm G}|^2\ell^2/4)$ knocks down higher shell contributions, the first Fourier coefficients $\beta_1$, $\delta_1$ and $\Delta_1$ will dictate the main effect of the effective periodic potential, through the coefficient
\begin{align}
   \xi^{(0,0)}_1\equiv\xi_1&= -\frac{\hbar\, e}{2m^*}\left(\frac{\beta_1\,G_+\,G_-}{|{\bm G}|^2} \right)-\frac{e^2}{2m^*}\left( \frac{\beta_1^2}{|{\bm G}|^2}\right)-\delta_1+\Delta_1=-\frac{\hbar\,e}{2m^*}\,\beta_1-\frac{3 e^2\,a_M^2}{32\,\pi^2 m^*}\,\beta_1^2-\delta_1+\Delta_1.
   \label{Sup:Xi1}
\end{align}
In the previous two expressions, $\beta_i$ has units $\Phi_0/A_M$ while  $\delta_i$ and $\Delta_i$ are expressed in meV. Introducing full units in Eq. \eqref{Sup:Xi1} we obtain
\begin{align}
    \xi_1 = \hbar\omega_c\left( -\frac{\bar\beta_1}{2}-\frac{\sqrt{3}\bar\beta_1^2}{8\,\pi}-\bar{\delta}_1\right)+\Delta_1=\hbar\omega_c\,\Omega_1+\Delta_1,
    \label{MagicAngleSupplemental}
\end{align}
where the dimensionless parameters 
\begin{equation}
    \bar{\beta}_1=\frac{1}{4\pi}\int_{UC}d^2r {\bm B}_{\text{eff}}({\bm r})\,e^{i{\bm G}\cdot{\bm r}}\qquad\text{and}\qquad \bar{\delta}_1 = \frac{\sqrt{3}}{32\pi}\int_{UC} d^2r \left( \sum_{i=x,y}\left[\partial_i {\bm n}\right]^2\right)e^{i{\bm G}\cdot{\bm r}}
\end{equation}
are the ones plotted in Fig. 1(d)-(e), with ${\bm G}$ in the first shell of reciprocal lattice vectors and the integration taken over the moiré unit cell.\\\\
The derivation of Eq. \eqref{Sup:EffectiveFourier2} in the previous section, from which Eq. \eqref{MagicAngleSupplemental} follows, was done for ${\bm B}_{\text{eff}}$ pointing in the $+z$--direction, corresponding to $\psi\in (0,180^{\circ})$ in the continuum model. As seen in Fig. 1(a)-(b) in the main text, the effective magnetic field and kinetic potential both have peaks at the $m$-points in the unit cell, which yields $\beta_1<0$ and $\delta_1<0$. This, together with the observation that $\overline{\beta}_1^2\ll\overline{\beta}_1$, guarantees that $\Omega_1$ is positive. For $\psi\in (180^{\circ},360^{\circ})$ the effective magnetic field changes sign, while the kinetic potential remains unchanged. In this case, $\beta_1>0$ and $\delta_1<0$, but the term $\sim aA_+$ in the adiabatic Hamiltonian will change sign as well. In this way we recover the expression for $\xi_1$ presented in the main text, Eq. (15), for general $\psi$. We also note that the adiabatic approximation breaks down at model parameter $\psi=180^{\circ}$ because all
components of the pseudospin field vanish at the XM and MX points (tunneling always vanishes at these points).
The direction of the layer pseudospins is reversed at these points upon crossing the $\psi=180^{\circ}$ line in model space, 
and the direction of the effective magnetic field changes as a result.  The Landau level approach 
correctly predicts the sign of the highest hole band Chern number on both sides of the $\psi=180^{\circ}$ line, as seen in Fig. 2 of the main text.
\begin{figure}[h!]
\centering
\includegraphics[width=0.95\linewidth]{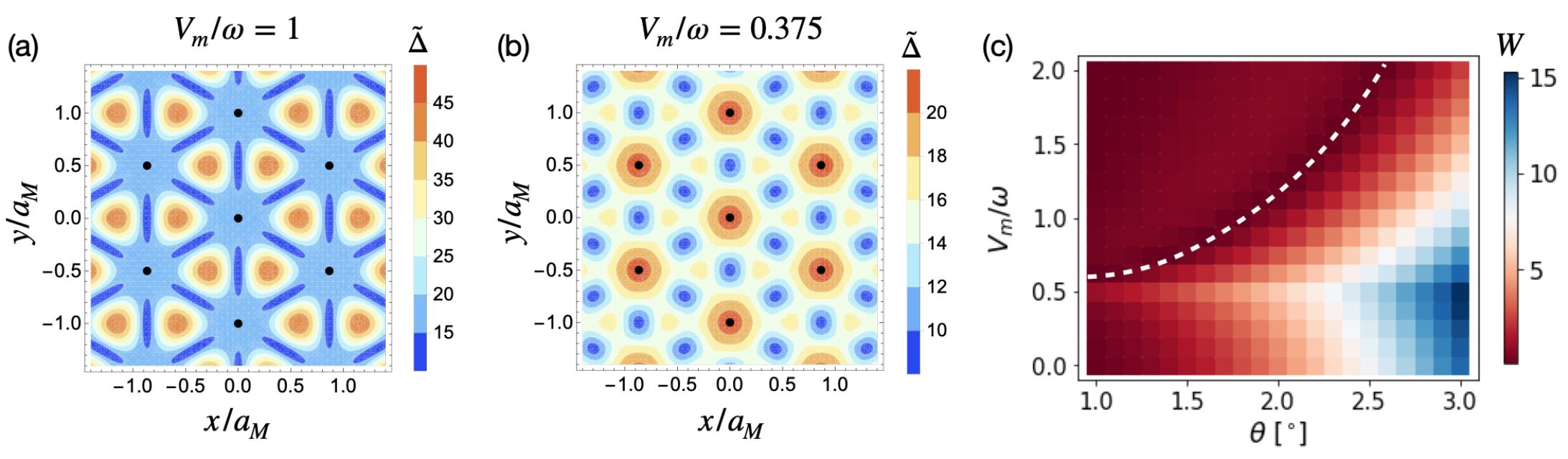}
\caption{(a)-(b) Spatial distribution of $\tilde{\Delta}$ for $V_m/\omega=1$ and $V_m/\omega=0.375$. The corresponding first shell Fourier coefficients are $\Delta_1=-2.339$ meV for (a) and $\Delta_1=0.867$ meV for (b). (c) Band width of the topmost moiré miniband from the continuum model Eq. \eqref{Sup:ContinuumModel} as a function of twist angle and the ratio $V_m/\omega$, for $\psi=100^{\circ}$. The evolution of the magic angle across the phase diagram is traced by the white dashed line. A magic angle emerges when the main peaks of $\tilde{\Delta}$ are located at the $\kappa$-points in the Wigner-Seitz cell.}
\label{fig:MagicLine}
\end{figure}

When our model is projected to the LLL and the effective potential is truncated to the first shell of reciprocal lattice vectors, the bandstructure can be obtained analytically (see the following section for the derivation). The resulting band width is proportional to $|\xi_1|$, hence the magic angle is determined by the vanishing of $\xi_1$, which is guaranteed provided $\Delta_1<0$. At the magic twist angle there is an effective cancellation between the local zero-point energy of the LLL and the effective Zeeman field. The exact cancellation of these two energy scales at the first-shell level is a manifestation of the property that the LLL is weakly sensitive to the high in-homogeneity of the effective potential.\\\\
The emergence of a magic angle in moiré TMD homobilayers depends crucially on the continuum model parameters, or equivalently in the shape of the pseudospin Skyrmion field ${\bm \Delta}$, which determines the sign of the coefficient $\Delta_1$. To illustrate this, in Fig. \ref{fig:MagicLine}(a)-(b) we show the spatial distribution of $\tilde{\Delta}$ for $V_m/\omega=1$ and $V_m/\omega=0.375$, with the main peaks located at $\kappa$ and $\gamma$ within the Wigner-Seitz cell, respectively. In Fig. \ref{fig:MagicLine} (c) we plot the band width of the topmost band obtained from the continuum model as a function of $\theta$ and the ratio $V_m/\omega$. There is a critical value of the ratio $V_m/\omega\approx 0.6$ after which a magic angle emerges and traces a {\it magic line} (white dashed line) in the phase diagram. In contrast, no magic angle behavior is observed below the critical value of $V_m/\omega$ in Fig. \ref{fig:MagicLine} (c). Below the critical ratio, the main peaks in $\tilde{\Delta}$ are located at $\gamma$, while above the critical ratio the main peaks move to $\kappa$. As pointed out in the main text, the latter case will lead to $\Delta_1<0$ and hence to the emergence of a magic angle. We expect that applying pressure to 
bilayers will increase the interlayer tunneling more strongly than it increases the moir\'e modulation 
potential, thereby decreasing $V_m/w$ \cite{FCI_Flatiron} and altering $\xi_1$.  
This observation motivates experiments that seek to tune the bilayers across phase transitions and 
theoretical work aimed at specific predictions in different filling factor regimes.\\\\
%
An alternative version to explain the cancellation of scales that gives rise to a magic angle can be obtained by directly defining the kinetic-momentum operators without separating the effective magnetic field into constant and moiré-periodic parts, $\Pi_{\alpha}=k_{\alpha}+\frac{e}{\hbar}\tilde{A}_{\alpha}$, which allows to rewrite the adiabatic Hamiltonian as
\begin{align}
        H&=-\frac{\hbar^2}{2m^*}\left(\Pi_x-i\,\Pi_y \right)  \left(\Pi_x+i\,\Pi_y \right) +\frac{e\,\hbar}{2m^*}B_{\text{eff}}-D+\tilde{\Delta}\nonumber \\
        &=-\frac{\hbar^2}{2m^*}\Pi_-\Pi_++\frac{e\,\hbar}{2m^*}B_{\text{eff}}-D+\tilde{\Delta}.  
\end{align}
A result first obtained by Aharonov and Casher for spin-$1/2$ charged particles in an arbitrary magnetic field \cite{aharonov1979ground} tells us that the Hamiltonian $\Pi_-\Pi_+$ has a manifold of exact zero energy ground states, whose wavefunctions are given by
\begin{align}
\psi({\bm r}) = f(z) \, e^{\phi({\bm r})}, \qquad\qquad 
-\nabla^2 \phi({\bm r}) = B({\bm r}).
\end{align}
The first-shell Fourier coefficient of the periodic function $e\,\hbar\,B_{\text{eff}}/2m^*-D+\tilde{\Delta}$ is given by
\begin{align}
  \xi_1^{\prime}=\hbar\omega_c\left( \frac{|\bar\beta_1|}{2} -\bar{\delta}_1\right)+\Delta_1.
\end{align}
If this coefficient vanishes, we are guaranteed an ideal flat band because the Hamiltonian reduces to $\Pi_-\Pi_+$. Note that this expression differs from Eq. \eqref{MagicAngleSupplemental} for the magic angle condition only by the term $\sim \overline{\beta}_1^2$, which is negligible since $\overline{\beta}_1^2\approx 0.047$ is an order of magnitude smaller than $\overline{\beta}_1$ and $\overline{\delta}_1$. We also note that the cancellation between energy scales leading to the magic angle is only exact when we truncate to the first shell of reciprocal lattice vectors. If further reciprocal lattice shells are considered, the band width will still reach a minimum but will not vanish, as we will elaborate in a future publication.\\\\
Finally, we discuss Landau-level mixing in the first-shell approximation. The Fourier coefficients of the matrix elements mixing the $n=0$ and $n=1$ Landau levels can be obtained from Eq. \eqref{GeneralFourierCoefficents} to be
\begin{align}
    \xi^{(1,0)}_1=\hbar\,\omega_c\left(-\frac{1}{2\sqrt{2}}\,\overline{\beta}_1+\frac{\sqrt{3}}{4\pi}\,\overline{\beta}_1-\frac{\sqrt{3}\,\overline{\beta}_1^2}{8\pi}-\overline{\delta}_1 \right)+\Delta_1.
\end{align}
The energy scale that determines mixing between the two lowest Landau levels is then
\begin{align}
    \eta_1=\left|6\,\xi^{(1,0)}_1\, \frac{G_0\ell}{\sqrt{2}}\,e^{-{\bm G}^2\ell^2/4}\right|= 6\,|\xi^{(1,0)}_1|\,\frac{\sqrt{2 \pi}}{3^{1/4}}e^{-\pi/\sqrt{3}} ,
\end{align}
where the factor of $6$ accounts for the number of reciprocal lattice vectors in the first shell. This is the quantity that is plotted in red in Fig. 3(c) of the main text. Interestingly, there is also a twist angle for which this Fourier coefficient will vanish and therefore LL-mixing vanishes as well, but this twist angle slightly differs from the magic angle. However, this indicates that LL-mixing will remain small at the magic angle.\\\\
Because at the magic angle our model becomes exactly the LLL from the homogeneous part of the effective magnetic field ${\bm B}_0$, when interactions are added the ground state for fractional fillings $\nu=1/m$, with $m$ an odd integer, will belong to the same universality class as the Laughlin state \cite{Laughlin}. If we perturbatively move away from the magic angle, the periodic potential will induce mixing between the ground state and excited states projected to the LLL (we ignore LL-mixing since it is weak). The first order intra-LL correction to the Laughlin-like state, denoted $\ket{\Psi_0^{(0)}}$, is given by
\begin{align}
\ket{\Psi_0^{(1)}}=\ket{\Psi_0^{(0)}}+\sum_{\bm G}\frac{\xi_1\braket{\Psi_{\bm G}^{(0)}|\overline{\rho}_{\bm G}|\Psi_0^{(0)}}}{E_0-E_{\bm G}}\,\ket{\Psi_{\bm G}^{(0)}}=\ket{\Psi_0^{(0)}}-\frac{\xi_1\,N\,\overline{S}({\bm G})}{\Delta({\bm G})}\sum_{\bm G}\ket{\Psi_{\bm G}^{(0)}},
\end{align}
where $\overline{S}({\bm G})=\braket{\Psi_0^{(0)}|\overline{\rho}^{\dagger}_{\bm G}\,\overline{\rho}_{\bm G} |\Psi_0^{(0)}}$ is the projected structure factor of the Laughlin state at momentum ${\bm G}$; $\Delta({\bm G})$ is the energy of the excitation corresponding to momentum ${\bm G}$, $N$ is the number of particles and the vectors ${\bm G}$ belong to the first shell of reciprocal lattice vectors. Close to the magic angle the coefficient $\xi_1$ is perturbatively small, hence the gap to excitations remains finite in the thermodynamic limit. Additionally, the structure factor should not display any sharp peaks in the Laughlin state, therefore the term $\xi_1 N\overline{S}/\Delta({\bm G})$ remains small for twist angles around the magic angle. In such region, the ground state will still be in the same universality class as the Laughlin state. This guarantees that our model supports FCI ground states.
\section*{Bandstructure of the lowest Landau level in a triangular periodic potential}
For a periodic potential with $C_3$ symmetry, the first six reciprocal lattice vectors should satisfy
\begin{align}
    \xi({\bm G}_1)=\xi({\bm G}_3)=\xi({\bm G}_5)=|\xi|e^{-i\,\chi},\qquad \xi({\bm G}_2)=\xi({\bm G}_4)=\xi({\bm G}_6)=|\xi|e^{i\,\chi},
\end{align}
for the Hamiltonian Eq. (12) to be Hermitian. In the following we will denote the norm of these six vectors as $G_0$. The effective potential resulting from the adiabatic approximation of TMD homobilayers has $C_6$ symmetry, which imposes $\chi=0$, however in this section we keep the analysis more general by taking $\chi$ to be arbitrary.\\\\
The solution of Eq. (12) in the main text is well known both in the square and hexagonal lattice cases \cite{Pfannkuche,Allan_QHE_Hexagonal,Claro,Yoshioka}, resulting in a set of $p$ sub-bands when the number of flux quanta per unit cell is $B_0\,A_{M}/\Phi_0=p/q$. In the following we will first focus on one flux quantum per unit cell $B_0=\Phi_0/A_M$. According to Eq. (12) in the main text, the diagonal matrix element only has contributions from terms in the potential whose corresponding reciprocal lattice vector $y$-component, $G_y$, vanishes
\begin{align}
    \varepsilon_0(X)=\braket{X|H|X}=-\frac{\hbar\omega_c}{2}-2|\xi|e^{-\frac{1}{4}G_0^2\ell^2}\cos(G_0\,X+\chi).
\end{align}
Similarly, the off-diagonal elements vanish unless $X'-X=\pm \delta X=\sqrt{3}G_0\ell^2/2$ and have contributions from reciprocal lattice vectors such that $G_y=\pm \delta \,X$
\begin{align}
    \varepsilon_{\pm}(X)=\braket{X\pm \delta X|H|X}=\braket{X|H|X\pm \delta X}= 2|\xi|e^{-\frac{1}{4}G_0^2\ell^2} \cos\left[ \frac{G_0}{4}(2X\pm\delta X)-\chi\right].
\end{align}
For one flux quantum per unit cell, the diagonal matrix elements are periodic under a translation under $n\,\delta X$, where $n$ is any integer
\begin{align}
    \braket{X+n\,\delta X|H|X+n\,\delta X}=2|\xi|e^{-\frac{1}{4}G_0^2\ell^2}\cos(G_0\,X+n G_0\delta X+\chi)=\varepsilon_0(X),
\end{align}
which follows because $n\,G_0\delta X=n \,2\pi$. The off-diagonal elements, in contrast, have a periodicity of $2\,n\, \delta X$,
\begin{align}
    \braket{X\pm\delta X+n\delta X|H|X+n\delta X}=2|\xi|e^{-\frac{1}{4}G_0^2\ell^2} \cos\left[ \frac{G_0}{4}(2X\pm\delta X+2\,n\,\delta X)-\chi\right].
\end{align}
This follows from the fact that for the triangular lattice, the $x$--components of the reciprocal lattice vectors ${\bm G}_2,{\bm G}_3,{\bm G}_5,{\bm G}_6$ are one half of the corresponding $x$--component of ${\bm G}_1$. This periodicity of the Hamiltonian allows to define a second wavevector, $k_x$, such that it can be written as tight--binding model
\begin{align}
    H(k_x,k_y)=\begin{pmatrix}
    \varepsilon_0(X)&\varepsilon_-(X)+e^{-i k_x(2\,\delta X)}\varepsilon_+(X)\\
    \varepsilon_-(X)+e^{i k_x(2\,\delta X)}\varepsilon_+(X)&\varepsilon_0(X)
    \end{pmatrix}.
\end{align}
The good quantum numbers $(k_x,k_y)$ take values $k_x \in [0,\pi/\delta X)$ and $k_y \in [0,\delta X/\ell^2)$. The resulting spectrum is given by 
\begin{align}
    E_{\pm}(k_x,k_y)=\varepsilon_0(k_y\,\ell^2)\pm \left[\varepsilon_-(k_y\,\ell^2)^2+\varepsilon_+(k_y\,\ell^2)^2-2\, \varepsilon_-(k_y\,\ell^2)\,\varepsilon_+(k_y\,\ell^2)\cos(2\,k_x\,\delta X)\right]^{1/2}.
    \label{TBSpectrum}
\end{align}
%
%
%
Plugging the expressions for $\varepsilon_0(x)$ and $\varepsilon_{\pm}(x)$ into Eq. \eqref{TBSpectrum}, the spectrum is
\begin{align}
    E_{\pm}(k_x,k_y)&=-\frac{\hbar \omega_c}{2}-2|\xi|\text{exp}\left(-\frac{\pi}{\sqrt{3}}\right)\left[ \cos(a_Mk_y+\chi)\pm\left[\cos^2\left(\frac{a_Mk_y}{2}-\frac{\pi}{2}-\chi\right)\right.\right.\nonumber\\
    &\left.\left.+\cos^2\left(\frac{a_Mk_y}{2}+\frac{\pi}{2}-\chi\right)-2\cos(\sqrt{3}a_Mk_x)\cos\left(\frac{a_Mk_y}{2}-\frac{\pi}{2}-\chi\right)\cos\left(\frac{a_Mk_y}{2}+\frac{\pi}{2}-\chi\right) \right]^{1/2} \right].
\end{align}
In particular for $\chi=0$, as in the case of the main text, we get
\begin{align}
    E_{\pm}(k_x,k_y)=-\frac{\hbar \omega_c}{2}-2|\xi|e^{-\pi/\sqrt{3}}\left[ \cos(a_Mk_y)\pm\left(2\sin^2\left( \frac{a_Mk_y}{2}\right) +2\cos(\sqrt{3}a_Mk_x)\sin^2\left(\frac{a_Mk_y}{2}\right)\right)^{1/2}\right].
\end{align}
The band width of this two--band spectrum is what we plot in Fig. (3) of the main text to compare with the continuum model band width.
%
\end{document}